\documentstyle[12pt]{report}
\begin{document}
\begin{center}
\Large
VALLEY METHOD VERSUS INSTANTON-INDUCED EFFECTIVE LAGRANGEAN UP TO
$(E/E_{\rm sphaleron})^{8/3}$.

\vskip 0.9 cm
I. Balitsky$^*$
\normalsize

\vskip 0.6 cm
Physics Deptartment, Penn State University

104 Davey Lab., University Park, PA 16802, USA

\vskip 0.6 cm
and

\vskip 0.6 cm
\Large
A. Sch\"afer
\normalsize

\vskip 0.6 cm
Institut f\"ur Theoretische Physik, Universit\"at Frankfurt,

Postfach 11 19 32, D-6000 Frankfurt

\vskip 0.9 cm
Abstract
\end{center}
We compare the two most popular approaches to the problem of
instanton-antiinstanton interaction at high energies - the valley
method and the effective-Lagrangian approach - and use them to
calculate the next-to-next-to-leading term in the expansion of
``holy grail'' function determining the cross section with baryon
number violation in the Standard Model.
\vskip 4 cm
{}~$^*$ On leave of absence from St. Petersburg Nuclear Physics
Institute, 188350 Gatchina, Russia
\newpage

\setcounter{chapter}{1}
\vskip 1 cm
\LARGE
1. Introduction
\normalsize
\vskip 0.6 cm

In the last three years there was a surge of interest in
instanton-induced processes leading to baryon number violation (BNV) in
the standard model. Although the very existence of instanton-induced
baryon (and lepton) number violation was known from the pioneering work
 of t'Hooft [1] these processes were considered to be of only academic
interest since the predicted probabilities all seemed to be of the
order exp($-16\pi^2/g_w^2)\sim 10^{-170}$. This scale is set by the
square of the Gamow factor corresponding
to instanton tunneling for a potential barrier of hight $E_{\rm
spha}\sim m_W/\alpha_W \sim 10$ TeV ($m_W$ is the $W$ boson mass and
the subscript $spha$ stands for
'sphaleron' which is the classical configuration leading over the top of the
barrier).

The situation changed, when Ringwald [2] suggested that the barrier
penetration probability could be strongly enhanced for collisions of
particles with energies comparable to the barrier height. It is even
hoped that BNV might be observable at SSC. Up to now this suggestion
could be  neither confirmed nor rejected
(although there are arguments both
pro [3,4] and contra[5-7])
since no technique is known to reliably calculate instanton-induced
processes at sphaleron energies.
What has been done (and what we continue to do in this contribution) is
to calculate these cross sections at small energies $\ll E_{\rm spha}$.
In doing so one hopes that some way can be found to extract from the
results also the high energy behaviour (similar to e.g. the summation of
leading logs in perturbation theory).
At small energies $E\ll E_{\rm sph}$ the BNV cross section turnes out
to be determined by the instanton-antiinstanton interaction at large
separations which is given by the well-known dipole-dipole formula [8].
Indeed, due to the optical theorem the BNV cross section is determined
by the imaginary part of the forward scattering amplitude in the
background of theinstanton-antiinstanton  ($I\bar I$) configuration.
 The relevant exponential term is
\begin{eqnarray}
\sigma_{BNV} &\sim& {\cal I}m \int d\rho_1d\rho_2dRdu~ \exp\left(
ER_0-{16\pi^2\over g^2} \right.
\nonumber \\
&&\left. +{32\pi^2\over
g^2R^6}(4(u\cdot R)^2-R^2)\rho_1^2\rho_2^2
 -\pi^2v^2(\rho_1^2+\rho_2^2)\right)
\end{eqnarray}
where $\rho_1$ and $\rho_2$ are the $I$ and $\bar I$ sizes, $R$ is the
separation and $u$ is the 4-vector determining the SU(2) matrix of relative
 $I\bar I$ orientation in isospin
space. The origin of the exponential terms in eq.(1) is the following:

\noindent (i)
$ER_0$ comes from the initial (and final) particles (the Euclidean
calculation of the $I\bar I$ contribution should be performed at
imaginary energy in order to obtain the necessary imaginary part, see
refs. [9,10]).

\noindent (ii)
The second term is twice the classical action of an instanton (which is
exactly the WKB suppression factor discussed above).

\noindent (iii)
The third term is the dipole-dipole $I\bar I$ interaction potential.

\noindent (iv)
The last term is the classical action due to the Higgs component of the
instanton (it is this term which makes the $\rho$ integrals convergent
unlike in QCD).

The integral in eq.(1.1) is dominated by the saddle point
\begin{equation}
\vec R_*=0~~,~~u_*^{\mu}~ \|~ R^{\mu} ~~,~~R_{0*}={
\epsilon^{1/3}\sqrt{6}\over
m_W} ~~,~~ \rho_{*1}=\rho_{*2}=\sqrt{{3\over 2}}
{\epsilon^{2/3}
\over m_W}
\end{equation}
such that with exponential accuracy
\begin{equation}
\sigma_{BNV} \sim ~ \exp\left( -{16\pi^2\over g_W^2}(1-{9\over 8}
\epsilon^{4/3})\right)
\end{equation}
where we have used the standard notation $\epsilon=E/E_0$,
$E_0=\sqrt{6}\pi m_W/ \alpha_W \sim$ $17$~TeV. This is the result of ref.
[2] (but with the correct numerical coefficient [9,11]) which lead to
so much enthusiasm since the Gamow factor is cancelled exactly at SSC
energies! Unfortunately eq.(1.3) is only valid at low energies $E\ll
E_{\rm sph}$ and has to be modified at higher energies. It is
convenient to introduce the so-called `holy grail' function
$F(\epsilon)$ determining the BNV cross section with exponential
accuracy
\begin{equation}
\sigma_{BNV} \sim ~ \exp\left( -{16\pi^2\over g_W^2}F(\epsilon)
\right)
\end{equation}
(the form of eq.(1.4) is fixed by dimensional considerations [9]). Then
eq.(1.3) gives the first two terms in the expansion of $F(\epsilon)$ for
small $\epsilon$. Up to now one additional term, proportional to
$\epsilon^2$ had been calculated and in this paper we determine the
fourth term
\begin{equation}
F(\epsilon)=1-{9\over 8}\epsilon^{4/3}+
{9\over 16}\epsilon^2-
{3\over 32}(4-3{m_H^2\over m_W^2})\epsilon^{8/3}\ln {1\over \epsilon}+
O(\epsilon^{8/3}\cdot const)
\end{equation}
where $m_H$ is the Higgs mass which we consider to be of order $m_W$.
The expansion in powers of $\epsilon^{2/3}$ reflects an expansion of
the $I\bar I$ interaction potential in powers of $\rho_*^2/R_*^2$ ( $=
\epsilon^{2/3}/4$,see
eq.(1.2). Of course, the main goal is to find the holy grail function at
large $\epsilon \ge 1$ and see whether it comes close to zero as
advocated in refs. [3,4] or is bounded from below by ${1\over 2}$ (the
one instanton action) as suggested in refs. [6,7]. In the former case
BNV might be observable at SSC while in the latter case it stays
exponentially small. However,
without a breakthrough in the calculations of
$F(\epsilon)$ at $\epsilon\sim 1$ one can only attack this problem by
calculating more terms at small $\epsilon$ and extrapolating
these results to higher and higher energies.

At small $\epsilon$ we face the familiar situation that the $I\bar I$
separation is much greater than the instanton sizes ($\rho_*^2/R_*^2 =
{1\over 4}\epsilon^{2/3}$, see eq.(1.2)).
There is a number of approaches to describe the instanton-antiinstanton
interaction at large separations and all were applied to calculate the
$\epsilon^{4/3}$ term in the expansion of $F(\epsilon)$ in eq.(1.5). It
was found first in ref.[2] by direct summation of $2\rightarrow N$
BNV-amplitudes in the instanton background. Soon after this it was
realized that the optical theorem relates the $\epsilon^{4/3}$ term to
the $I\bar I$ dipole-dipole interaction at large distances [11]. The
connection between these two calculations can be explained most easily
using the instanton-induced effective Lagrangean [11,12].
\begin{eqnarray}
L_{eff}&&=\int dx~\int {d\rho\over \rho^5}~ \int
du~d(\rho)~\exp(-2\pi^2\rho^2\bar\phi(x) \phi(x))
 \\
&&\left \{ \exp\left(
{2\pi^2i\over g} \rho^2~ Tr\{ \sigma_{\alpha}\bar \sigma_{\beta}~G_{\alpha
 \beta}(x)\}\right )~+~ \exp\left (
{2\pi^2i\over g} \rho^2~ Tr\{ u \bar \sigma_{\alpha}
\sigma_{\beta}\bar u ~G_{\alpha \beta}(x)\}\right )\right\}
\nonumber
\end{eqnarray}
where the first term in braces correspond to instanton and the second
to $\bar{I}$.Here $\rho$ and $x$ are size and center of would-be
(anti)instanton, $u\equiv u_{\mu}\sigma_{\mu}$ (respectively $\bar
u\equiv u_{\mu}\bar \sigma_{\mu}$) is the matrix of relative $I\bar I$
orientation, and $d(\rho)$ is the usual instanton density
given by the $exp(-{8\pi^2\over g_W^2})$ times quantum determinant in
the instanton background. We use the four-dimensional Pauli matrices
$\sigma_{\mu}=(1,-i\vec\sigma)$,
$\bar \sigma_{\mu}=(1,i\vec\sigma)$
related to the t'Hooft $\eta$-symbols by
\begin{equation}
\sigma_{\mu}\bar
\sigma_{\nu}~=~\delta_{\mu\nu}+i\bar{\eta}^a_{\mu\nu}\tau^a~~
,~~\bar\sigma_{\mu}
\sigma_{\nu}~=~\delta_{\mu\nu}+i\eta^a_{\mu\nu}\tau^a
\end{equation}
In principle, the effective Lagrangean contains an infinite series of
operators with growing dimension (see section 4). However, up to order
$\epsilon^{8/3}\ln \epsilon$ we shall need only the terms given in
eq.(1.6).

Using this Lagrangean, the instanton (in the singular gauge) can be
depicted as a local vertex from which an arbitrary number of W bosons
and Higgs particles can emerge. Every W boson provides a factor
$\rho^2/g$ and every Higgs a factor $\rho$ (or $\rho^2v $ after the
usual shift $\phi\rightarrow \phi+v/\sqrt{2}$).
After contracting the W's emitted by the effective vertex (6) with the
usual perturbative vertices
the Green functions in the instanton background (and the instanton
field itself) are obtained as power series in $\rho^2$.
This is illustrated in Fig.1a for the classical field and in Fig.1b for
the Green function.
(Gauge fields are depicted as wavy lines, Higgs fields as plain
ones.)
Note that due to cancellations between the
factors $g$ and $\rho^2/g$ each loop adds a factor $\rho^2$ rather than
$g^2$. The dipole-dipole $I\bar I$ interaction is then given by the sum
of diagrams shown in Fig.2. This sum corresponds to the
exponentiation of the first non-trivial diagram which can be easily
calculated giving the dipole-dipole interaction potential
\begin{eqnarray}
&&{2\pi i\over g} \rho_1^2~  Tr
\left\{ u \bar \sigma_{\alpha} \sigma_{\beta} \bar
u {\tau^a\over 2}\right\}~
{2\pi i\over g} \rho_2^2 Tr
\left\{ \sigma_{\mu} \bar \sigma_{\nu}
{\tau^b\over 2}\right\} ~\overline{G_{\alpha\beta}^a(R) G}_{\mu\nu}^b(0)
\nonumber\\
&&=~{32\pi^2\over g^2} ~{\rho_1^2\rho_2^2\over R^4}~\left(4{(u\cdot R)^2\over
R^2}-1+O(m_W^2R^2)\right)
\end{eqnarray}
The BNV cross section in eq.(1.1) is obtained by continuating to
Minkowski energies and taking the imaginary part in the saddle point
(1.2). On the other hand, this imaginary part can be taken at the level
of the diagrams in Fig.2. Since in this order in $\epsilon$ the only
effect of incoming particles is to provide energy we obtain exactly the
same series of $2\rightarrow N$ cross sections summed up in ref.[2].

The next-to-leading term in the expansion (1.5) $\sim\epsilon^2$ was
first calculated in ref.[10] with the valley method [13-15]. Using the
approximate conformal invariance of the pure gauge theory at tree level
it could be shown that the gauge part of the $I\bar I$ interaction
potential depends only on the conformal parameter
\begin{equation}
\zeta~=~ {R^2+\rho_1^2+\rho_2^2 \over \rho_1\rho_2}
\end{equation}
Since $U_{\rm int}^{\rm gauge}(\zeta)$ is expanded in powers of
$\zeta^2$ the dipole-dipole term $(32\pi^2/g^2\zeta^2)$
$(4\cos ^2\phi-1)$
contributes to the next-to-leading power in $\rho^2/R^2$. After adding
the simple part of $U_{\rm int}$ which is due to the Higgs field it gives the
$\epsilon^2$ term in the expansion of $F(\epsilon)$. Later this result
was reproduced by direct calculations of the amplitudes in the
instanton background [16-18] which in the language of an effective
Lagrangean correspond to the diagrams shown in Fig.3.
The first of the diagrams (Fig.3a) gives the term proportional to
$\rho^6/(g^2R^6)$ in the expansion of the gauge part of $U_{\rm int}$.
It coincides with the one obtained from the conformal valley. The
second diagram (Fig.3b) gives the mass correction to the dipole-dipole
term which is of order $(m^2\rho^4/(g^2R^6)$ (see eq.(1.7)) and the last
diagram describes the part of the $I\bar I$ potential which is due to
Higgs exchange and proportional to $v^2\rho^4/R^2$. For the saddle point
(2)
all these diagrams
contribute to the order $\epsilon^2$ and their sum reproduces the
valley result.

As to the last term in eq.(1.5) which is proportional to
$\epsilon^{8/3}\ln \epsilon$, there exists a calculation [19] for the
gauge part of $U_{\rm int}$ ($\sim (\rho^8/g^2R^8 \ln(R/\rho)$)
using the effective Lagrangean. In this calculation many two-loop
diagrams had to be summed. All of them are listed in section 5,
Fig. 4a shows typical examples.
It should be emphasized, that unlike for lower orders (up to
$\rho^6/R^6$) in this order the effective Lagrangean result [19]
and the
conformal valley result [14,15]
for the gauge part of $U_{\rm int}$ do not
coincide.
The reason is that in the conformal valley approach,
starting at order $\rho^8/R^8$, the contribution from
the gauge part of $U_{\rm int}$ depends on the specific choice for the
valley, a dependence which is canceled by the Higgs part of $U_{\rm
int}$ (see below). Similarly for the effective Lagrangean
approach,
the gauge part of $U_{\rm int}$
should be supplemented by a $\epsilon^{8/3}$ contribution from diagrams
containing Higgs bosons (an example is shown in Fig. 4b) and from mass
corrections to the diagrams in Fig.3a.
If all contributions are properly included both approaches give the
same $\epsilon^{8/3}\ln \epsilon$ term.

It is very important that the term proportional to $\epsilon^{8/3}$ is
the last one in equation (1.5) which is known to be determined
exclusively by the $I\bar I$ interaction. Starting at order
$\epsilon^{10/3}$ the situation becomes unclear. From this order on
hard-hard corrections due to exchanges between the initial or final
particles enter the game. It was argued in ref.[20]
(see, however, the recent ref. [21])
that these
corrections could exponentiate such that $F(\epsilon)$ would depend on
the initial and final states and not on $U_{\rm int}$ alone.
Then $F(\epsilon)$ could depend on the specific BNV process.

For the same reason the $\epsilon^{8/3}$ term is the last one obtainable
by analytic continuation of an Euclidian calculation for the forward
scattering amplitude in the $I\bar I$ background. The reason is that
the intermediate states with and without BNV
lead to two discontinuities (in energy) of the hard-hard corrections
(see Fig. 5a and 5b). Both contribute to the imaginary part of the
analytically continued eucledian diagram, but only Fig.5a contributes
to the BNV cross section (a detailed discussion can be found in
ref.[22]). Up to order $\epsilon^{8/3}$ the corresponding diagrams
contain no hard-hard corrections and thus have only the one
discontinuity (of the type shown in Fig.5a) which can be obtained from
the imaginary part of the continued Euclidian diagrams.

\setcounter{chapter}{2}
\setcounter{equation}{0}
\vskip 1 cm
\LARGE
2. Valley method
\normalsize
\vskip 0.6 cm

The saddle-point gaussian approximation is the usual technique to
calculate functional integrals in weak coupling theories. The valley
method [13] constitute a generalization to cases in which
physically relevant `approximate solutions' can be given.
The case of an instanton-antiinstanton pair at large separation is a
typical example. To illustrate the idea we consider quantum mechanics
in a double-well potential with the instanton being the simple kink
\begin{equation}
\phi_{I}~=~{1\over 2} ({\rm th}{\alpha-t \over 2}~+~1)  ~~~~~~~
\phi_{\bar I}~=~{1\over 2} ({\rm th}{t+\alpha \over 2}~+~1)
\end{equation}
describing the tunneling between the two minima. We want to calculate
the non-perturbative part of the vacuum energy
\begin{equation}
Z~=~N^{-1}~\int {\cal D}\Phi ~ \exp \left\{ -{1\over g^2}~\int ~dt~
{1\over 2} [(\stackrel{.}{\Phi} )^2 +\Phi^2(1-\Phi^2)] \right\}
\end{equation}
which at small $g^2$ is dominated by the $I\bar I$ configurations
[24].
For infinitely large separations the $I \bar I$ configuration is just
the sum of two kinks. It obeys the classical field equations and
possesses two zero modes corresponding to the independent translations
of both instantons. The zero modes can be rediagonalized in such a way
that one of them describes the trivial translations of the complete
$I\bar I$ configuration and the other changes in the instanton
separation. For large but finite separations (as compared to the
instanton size which we chose as 1 for this toy model) the second one
becomes a quasizero mode: the action varies slowly along this direction
in functional space but grows rapidly in orthogonal directions, which
correspond to changes in the instanton profile and are associated with
normal modes. As a landscape in functional space this looks like a
steep canyon with the course of the valley corresponding to the
quasizero mode (see Fig.6).

In order to integrate over the field configurations close to the $I
\bar I$ valley one has to perform the following steps:
(i) determine the course of the valley in the functional space,~
(ii) perform the Gaussian integrations in the directions orthogonal to
this valley, and (iii) carry out the final integration along the valley.
In step (i) one determines the valley as the trajectory in functional
space
which
minimizes the action. For any direction orthogonal to the valley the
constraint
\begin{equation}
(\Phi-\Phi_v, \omega(\alpha){\partial\Phi_v\over \partial\alpha})=0
\end{equation}
must be fulfilled, where $(f,g)$ denotes the usual scalar product of
functions $\int dt f(t)g(t)$ and $\omega(\alpha,t)$ is a suitable
weight function. Applying the standard technique of Lagrange
multipliers this leads to the following classical constraints (the
`valley equations'[13])
\begin{equation}
\left. {\delta S\over \delta \Phi(t)}
\right|_{\Phi=\Phi_v}~=~\chi(\alpha)
\omega (\alpha,t) {\partial\Phi_v(\alpha,t)\over \partial\alpha}
\end{equation}
where $\chi(\alpha)$ is the Lagrange multiplier. This equation has to
be solved with the boundary condition that for infinite $\alpha$
$\Phi_v$ approaches the field of an infinitely separated $I\bar I$
pair.
\begin{equation}
\Phi_v(\alpha,t) ~\rightarrow~ \Phi_I(t-\alpha)~+~ \Phi_{\bar
I}(t+\alpha) ~-~1
\end{equation}
As the valley action increases monotonously with $\alpha$
\begin{equation}
{\partial S_v(\Phi(\alpha))
\over  \partial \alpha} ~=~ \chi(\alpha)~ \left(
{\partial \Phi_v\over \partial \alpha}, \omega
{\partial \Phi_v\over \partial \alpha} \right) ~\ge~ 0
\end{equation}
we can conclude that $\alpha=0$ corresponds to the classical
perturbative vacuum and that $S$ reaches a finite value for large
$\alpha$ only if $\chi(\alpha) \rightarrow 0$, i.e. for a classical
solution (see eq. (2.4)).
Generally speaking the valley always interpolates between two classical
solutions, in our case the $I\bar I$ configuration and the vacuum.

In order to integrate over the orthogonal Gaussian modes (step (ii)) we
use the standard Faddeev-Popov trick and insert
\begin{equation}
1~=~ -\int d\alpha~ \delta(\Phi-\Phi_v,\omega \Phi_v')~
\{ (\Phi_v,\omega\Phi_v')-
(\Phi-\Phi_v, (\omega \Phi_v')')\}
\end{equation}
with $\Phi_v'\equiv d\Phi_v/d\alpha$. (Strictly speaking an additional
$\delta$-function, $\delta((\Phi-\Phi_v,d\Phi_v/dt))$
is needed to exclude total translations from
the integral (zero mode). This adds some technical complexity without
changing the arguments [13].)
Then we shift $\Phi$ to $\Phi+\Phi_v$ and expand the action in powers
of $\Phi$
\begin{equation}
S(\Phi+\Phi_v)~=~S(\Phi_v)+(\Phi,J_v)+{1\over 2} (\Phi,
\Box_v\Phi)+O(\Phi^3)
\end{equation}
where $J_v=\delta S/\delta \Phi_v$ and
$\Box_v=-\partial^2+1-6\Phi_v+6\Phi_v^2$ is the operator of the second
derivative of the action.

Now comes the central point: the linear term $(\Phi,J_v)$ in the
expansion (2.8) vanishes due to the factor $\delta(\Phi,\Phi_v')$ in
the integrand and the valley equation (2.3).

Thus $\Phi_v$ enters the functional integral like a classical solution
(for which $J=0$).
\begin{equation}
N^{-1}\int~d\alpha~ (\Phi_v,\omega\Phi_v')~e^{-{1\over g^2} S(\Phi_v)}
{}~\int{\cal D}\Phi~ \delta(\Phi, \omega\Phi_v')~e^{-{1\over
2g^2}(\Phi,\Box_v \Phi)}~(1+O(g^2))
\end{equation}
All effects $\sim 1/g^2$ originate from the classical action. Quantum
corrections come from the terms of order $\Phi^3$ in eq.(2.8) and from
the collective coordinate Jacobian (2.7).

Finally in step (iii) the explicite integration over the valley
parameter $\alpha$ must be performed. In this case it gives the
non-perturbative part of the vacuum energy, see ref. [13].

The crucial point of the whole procedure is the vanishing of the linear
term $(\Phi,J)$, which will occure for any weight function
$\omega(\alpha,t)$. Hence, at first sight, any valley starting from
infinitely separated instantons and antiinstantons seems appropriate.
The fact that some choices are worse than others shows up in the size
of the quantum corrections. A `good' valley should minimize them. The
standard recipe to find such a valley is the following: start from
infinitely separated instantons and follow the direction of the
negative quasizero mode of the operator $\Box_v$ (see the discussion in
ref.[22]).
\begin{eqnarray}
\Phi_v& ~~\stackrel{\alpha\rightarrow \infty}
{\longrightarrow}~~&
{1\over 2} {\rm th}{t+\alpha\over 2} ~-~
{1\over 2} {\rm th}{t-\alpha\over 2} \nonumber \\
\Phi_v'& ~~\stackrel{\alpha\rightarrow \infty}
{\longrightarrow}~~&
\Phi_- ~\sim~ {\rm ch}^{-2}{t+\alpha\over 2} ~+~
{\rm ch}^{-2}{t-\alpha\over 2}
\end{eqnarray}
All valley satisfying eq.(2.10) are appropriate to calculate the
nonperturbative part of the vacuum energy. (Of course, the final answer
obtained after integrating over the valley parameter $\alpha$ in (2.9)
is the same for all valleys.) The simplest choice for such a valley is
the sum of the kinks
\begin{equation}
\Phi_v ~=~
{1\over 2} {\rm th}{t+\alpha\over 2} ~-~
{1\over 2} {\rm th}{t-\alpha\over 2}
\end{equation}
which trivially satisfies the condition (2.10) and obeys the valley
equation (2.4) with the weight function
\begin{equation}
\omega(\alpha,t)~=~ {1\over 4}{ e^{\alpha} {\rm sh}
(\alpha)  \over {\rm ch}(t)
{}~{\rm ch}(\alpha) ~+1}
\end{equation}
The corresponding Lagrange multiplier is
\begin{equation}
\chi(\alpha)={12\over \zeta^2} ~~~,~~~ \zeta=e^{\alpha}
\end{equation}
and the valley action equals
\begin{equation}
S_v\equiv S(\Phi_v)=~{6\zeta^2-14\over (\zeta-1/\zeta)^2}~-~ {17 \over
3}~+~ \left[ {(5/\zeta~-~\zeta)(\zeta+1/\zeta)^2\over
(\zeta-1/\zeta)^3}~+1\right] \ln \zeta
\end{equation}
For $\alpha=0,~\zeta=1$ this expression gives zero
(perturbative vacuum) and with increasing $\alpha$ it approaches 1/3
which is twice the instanton action. Here $1/\zeta$ is the
small parameter corresponding to an expansion of the $I\bar I$
interaction at large separations.
(The conformal transformation will turn $\zeta$
into the expression (1.9).)
The leading terms in the expansion of the valley action are
\begin{equation}
S_v~=~ {1\over 3} ~-~ {2\over \zeta} ~+~ {12\over \zeta^4} \ln \zeta
{}~+~ ...
\end{equation}
In the integral (2.9) $\alpha$ is typically of the order $\ln (-g^2)$
(The sign of $g^2$ has to be changed in order to extract the
nonperturbative part of the partition function (2.9), see ref.[23].)
such that the third term $\sim \zeta^{-4}\ln \zeta$ (and higher ones)
in eq.(2.14) mix with the quantum corrections $O(g^2)$.
If there were an
extra parameter dominating $\zeta$ (e.g. the energy for the calculation
of the correlator $<\Phi(E)\Phi(-E)>$) the expansions in $g^2$ and
$\zeta$ would be independent. The latter is the case for BNV processes
where $\zeta \sim (E/E_{\rm sph})^{2/3}$.

If one is only interested in the first terms of the expansion (2.15),
say up to order $\zeta^{-2k}$, it is sufficient to fulfill the valley
equation (2.4) up to the order $\zeta^{-k-1}$.
\begin{equation}
J_v ~=~ \chi \omega {\partial \Phi_v\over \partial \alpha} ~+~
O(\zeta^{-k-1})
\end{equation}
To show this let us estimate the contribution of the linear term
of equation (2.8), namely
$(\Phi,J_v)$, to the action $S_v$ in the exponent of (2.9).
We obtain
\begin{eqnarray}
e^{-{S_v\over g^2}}~ \int {\cal D}\Phi~ \delta(\Phi,\omega \Phi_v')~
e^{-{1\over g^2}(\Phi,J_v)-{1\over 2g^2}(\Phi,\Box_v\Phi)} ~=~
\nonumber \\
e^{-{S_v\over g^2}+{1\over 2 g^2} (J_v,GJ_v)}~
\int {\cal D}\Phi~ \delta(\Phi,\omega \Phi_v')~
e^{-{1\over 2g^2}(\Phi,\Box_v\Phi)}
\end{eqnarray}
where  $G$ is the Green function of the operator $\Box_v$ with the
constraint
\begin{eqnarray}
G(t_1,t_2)&=&  N^{-1}  \int {\cal D}\Phi \delta(\Phi,\omega \Phi_v')
{1\over g^2}~\Phi(t_1) \Phi(t_2)
e^{-{1\over 2g^2}(\Phi,\Box_v\Phi)} =
(t_1|{1\over \Box_v}|t_2)
\nonumber \\
&&-~
(t_1|{1\over \Box_v}|\omega \Phi_v')~
(\omega\Phi_v'|{1\over \Box_v}|t_2)~
(\omega \Phi_v'|{1\over \Box_v}|\omega \Phi_v')^{-1}
\end{eqnarray}
(Here we have used the Schwinger notation for Green functions
$G(x,y)=(x|{1\ over\Box}|y)$,$(x|y)=\delta^4(x-y)$). Now
it is easy to see that the additional term in the exponent has the
order $\zeta^{-2k}$ since
\begin{equation}
(J_v,GJ_v)=((J-\chi\omega \Phi_v'),G
(J-\chi\omega \Phi_v')) \sim  {1\over \lambda_-}
((J-\chi\omega \Phi_v'),\Phi_-)^2 \sim \zeta^{-2k}
\end{equation}
and the eigenvalue corresponding to the negative quasizero mode is
$\sim \zeta^{-2}$ (see Ref.[13]). In practice it is often more
convenient to check directly the condition (2.19) for a given valley
than to verify the valley equation (2.16).
Starting from the double-well valley (2.11) it is easy to construct
a suitable valley for
massless gauge theories (QCD). Due to the conformal invariance of QCD
at the tree level it is possible to construct a whole family of $I\bar
I$ configurations with finite separattion from a spherically symmetric
configurations with separation zero [14]. It is known, that for this
spherical ansatz  (and for collinear gauge orientations) QCD is
equivalent at tree level to ordinary double-well quantum mechanics
(specified by eq. (2.2)). For
\begin{equation}
A_{\mu}(x) ~=~ -{i\over g}~ (\sigma_{\mu} \bar x-x_{\mu}) ~x^{-2}~
\Phi(t,\alpha)
\end{equation}
with $t=\ln x^2/\rho^2$ the QCD action coincides with the simple
quantum mechanical expression (2.2) up to an overall factor
$48\pi^2/g^2$. Using the quantum-mechanical valley (2.11)
we obtain thus the gauge field in the form
\begin{equation}
A_{\mu}(x)^v ~=~ -{i\over g}~ (\sigma_{\mu} \bar x-x_{\mu}) ~
\left( {\rho^2/\zeta \over x^2+\rho^2/\zeta}-
{\rho^2 \zeta \over x^2+\rho^2 \zeta} \right)
\end{equation}
which coincides with the sum of one instanton field in the regular
gauge with radius $\rho \sqrt{\zeta}$ and one antiinstanton field in
the singular gauge with radius $\rho/\sqrt{\zeta}$ up to a gauge
transformation with the matrix $\bar x/ \sqrt{x^2}$. This gauge field
obeys the valley equation
\begin{equation}
\left. {\delta S \over \delta A_{\mu}(x)}  \right|_{A=A_v}
 ~=~ \chi(\zeta)
\omega(x,\zeta) P^{\bot}_{\mu\nu}~ \zeta~ {dA_v^{\nu}(x,\zeta)\over
d\zeta}
\end{equation}
where $P_{\mu\nu}^{\bot}=\delta_{\mu\nu}-{\cal D}_{v,\mu}
(1/{\cal D}_v^2)
{\cal D}_{v,\nu}$
is a projector ensuring the decoupling of the gauge and non-gauge
constraints [14,15] (recall that the valley equation is a constraint
classical equation). The action of this projector on fields of the type
(2.20) is trivial since ${\cal D}_{\mu}A_{\mu}=0$. Therefore we shall
omit $P^{\bot}_{\mu\nu}$ in what follows (cf. refs. [14,15]). The
weightfunction $\omega(x,\zeta)$ is simply (2.11) taken at $t=\ln
x^2/\rho^2$.
To obtain the $I\bar I$ valley configuration for arbitrary sizes
$\rho_1,~\rho_2$ and separations $R$ one has to perform the translation
$x\rightarrow x-x_0$, the inversion
\begin{equation}
(x-a)_{\mu} \rightarrow {r^2\over (x-a)^2}~ (x-a)_{\mu}
\end{equation}
and a gauge transformation with the matrix $x(\bar x -\bar
R)R/\sqrt{x^2(x-R)^2R^2}$. The parameters $\rho_1,~\rho_2$ and $R$ are
then related to those of the initial configuration (2.21) and of the
canonical transformation (2.23) as follows.
$$
\rho_1~=~ {r^2\rho \sqrt{\zeta} \over (x_0-a)^2+\rho^2 \zeta} ~~~~,~~~~
\rho_2~=~ {r^2\rho/\sqrt{\zeta} \over (x_0-a)^2+\rho^2/\zeta} $$
\begin{equation}
R ~=~ (x_0-a)~ \left( {r^2\over (x_0-a)^2+\rho^2/\zeta}-
{r^2\over (x_0-a)^2+\rho^2 \zeta} \right)
\end{equation}
After some algebra one obtains the final answer for the gauge valley in
the form
\begin{equation}
A^v_{\mu}~=~ A^I_{\mu}~+~ A^{\bar I}_{\mu} ~+~ B_{\mu}
\end{equation}
where
\begin{equation}
A_{\mu}^I=-{i\over g} {\sigma_{\mu}\bar x -x_{\mu} \over x^4 \Pi_1}
\rho_1^2 ~~~,~~~
A_{\mu}^{\bar I}=-{i\over g} {R(\bar \sigma_{\mu} (x-R)
-(x-R)_{\mu})\bar R \over R^2 (x-R)^4 \Pi_2}
\rho_2^2 ~~~,~~~
\end{equation}
\begin{eqnarray}
&&B_{\mu}= {i\over g} ~{\rho_1\rho_2\over \zeta}~
\left\{  {x(\bar x-\bar R)\sigma_{\mu}
\bar x \over x^4 (x-R)^2 \Pi_1}-
{R(\bar x-\bar R)\sigma_{\mu}
\bar x (x-R)\bar R\over R^2x^2 (x-R)^4 \Pi_2} \right. \nonumber \\
 &&+{\sigma_{\mu}
\bar R\over x^2 (x-R)^2 \Pi_1\Pi_2} ~+~
\rho_1^2 \left( 1-{\rho_2\over \zeta\rho_1} \right)
{\sigma_{\mu}
\bar x\over x^4 (x-R)^2 \Pi_1\Pi_2} \nonumber \\
 &+&\rho_2^2 \left( 1-{\rho_1\over \zeta\rho_2} \right)
{R \bar \sigma_{\mu} (x-R)
\bar R\over R^2x^2 (x-R)^4 \Pi_1\Pi_2}
\nonumber \\
&-&
\left. {\rho_1\rho_2\over \zeta}~
{R (\bar x-\bar R) \sigma_{\mu} \bar x
\over x^4 (x-R)^4 \Pi_1\Pi_2} ~-~(trace)\right\}
\end{eqnarray}
where $\Pi_1=1+\rho_1^2/x^2$ and $\Pi_2=1+\rho_2^2/(x-R)^2$.
'$O-(trace)$'
 means the traceless part of $O$. (Strictly speaking one has to
subtract ${1\over 2} {\rm Tr}O$.)
Thus the valley field is a sum of instanton and antiinstanton (in a
singular gauge) with relative orientation collinear with $R$ (the
maximal attractive orientation) plus a small additional field
proportional to $1/\zeta$.
(The form of the valley used in ref.[22] differs from (2.25) by a
gauge transformation with the matrix $x(x-a) (x-b) (x-R)/ \sqrt{
x^2(x-a)^2(x-b)^2(x-R)^2}$, where $b=R+(x_0-a)r^2/(x_0-a)^2$.
The valley of ref.[15] is obtained by a slightly more complicated
gauge transformation.)

The valley equation for this configuration has the usual form
\begin{equation}
-{\cal D}_{\mu}G_{\mu\alpha} ~=~ 2\chi(\zeta)\omega(x,\zeta)\zeta~
{\partial A_{\alpha}\over \partial \zeta}
\end{equation}
where [16]
\begin{equation}
\omega(x,\zeta) ~=~ {\zeta^2-1\over (x^2+\rho_1^2)^2\rho_1^{-2}
+ ((x-R)^2+\rho_2^2)^2\rho_2^{-2}} ~~~~.
\end{equation}
$\chi(\zeta)$ is given by eq.(2.13) and $\zeta$ depends on the new variables
$\rho_1,~\rho_2$, and $R$ according to
\begin{equation}
\zeta ~=~ {R^2+\rho_1^2+\rho_2^2 \over 2\rho_1\rho_2}
{}~+~\sqrt{\left( {R^2+\rho_1^2+\rho_2^2 \over 2\rho_1\rho_2}\right)^2-1}
{}~+~ \sim
{R^2+\rho_1^2+\rho_2^2 \over \rho_1\rho_2}
\end{equation}
The classical action of this configuration coincides, of course, with
eq.(2.11)
\begin{eqnarray}
S~&=&~{48\pi^2\over g^2}~\left\{ {6\zeta^2-14\over
(\zeta-1/\zeta)^2}~-~ {17 \over
3}~+~ \left[ {(5/\zeta~-~\zeta)(\zeta+1/\zeta)^2\over
(\zeta-1/\zeta)^3}~+1\right] \ln \zeta \right\}  \nonumber \\
&&\sim {16\pi^2\over g^2}~
\left(1-{6\over \zeta^2}+{36\over \zeta^4}\ln \zeta + ...\right)
\end{eqnarray}
where the dots stand for non-logarithmic contributions $\sim \zeta^{-4}$
as well as for other higher terms. It is worth noting that the argument
of the  running coupling constant $g(\mu)$ in (2.31) could be taken as
 $\mu=\rho_1\rho_2$
with our accuracy (since at large $R$ it should reproduce
$8\pi^2/g^2(\rho_1)~+~8\pi^2/g^2(\rho_2)$ corresponding to independent
instantons).

\setcounter{chapter}{3}
\setcounter{equation}{0}
\vskip 1 cm
\LARGE
3. The $\epsilon^{8/3}\ln \epsilon$ term for gauge theories with
Higgs field in the valley approach
\normalsize
\vskip 0.6 cm

First let us reproduce the $\epsilon^2$ term in the expansion of the
holy grail function (1.5). To this end we calculate the BNV part of the
forward scattering amplitude of $W$ bosons
\begin{equation}
A(p,k)~=~ N^{-1} ~\int {\cal D}A~{\cal D}\phi
{}~{\cal D}\bar \phi~{\cal D}\psi
{}~{\cal D}\bar \psi~e^{-S}~ A_{\mu}^a(p)A_{\nu}^b(k)
A_{\mu}^a(-p)A_{\nu}^b(-k)
\end{equation}
by the valley method ($A_{\mu}(p)\equiv \int dx A_{\mu}(x)$exp($ipx$)).
We shall consider the conventional model without hypercharge
\begin{equation}
L={1\over 4} G_{\mu\nu}^aG_{\mu\nu}^a-\sum_{k=1}^{12} \bar \psi_k
i \bar {\not \nabla} \psi_k+ \left| \nabla \phi\right|^2 +
\lambda(\left|\phi\right|^2-v^2/2)^2
\end{equation}
As we shall see below we can neglect particle masses at order $\epsilon^2$
and use the $I\bar I$ valley in the form (2.26).
In addition we introduce
\begin{equation}
\phi_v=\phi_1 \phi_2 {ve \over \sqrt{2}}~~~,~~~
\bar \phi_v=\phi_1 \phi_2 {v\bar e \over \sqrt{2}}~~~,~~~
\phi_1 ={1 \over \sqrt{\Pi_1}}~~~,~~~
\phi_2 ={1 \over \sqrt{\Pi_2}}
\end{equation}
where $e$ is the unit vector which we chose to be
$1\choose 0$ (so $\bar
e=(1,0)$). (Recall that
$\phi_1 {ve \over \sqrt{2}}$ is the Higgs component of the instanton
satisfying the equation $\nabla_I^2\phi_1=0$
and similarly for $\bar I$.) The
fermion component of the valley could also be taken into account, e.g.
as the product of zero modes corresponding to $I$ and $\bar I$ (cf.
ref.[24]). However, to calculate $F(\epsilon)$ this is not necessary as
the fermions affect only the preexponential factor.

The calculation of the amplitude (3.1) with the valley method proceeds
as  for our example, the double-well vacuum energy.  We insert the
necessary $\delta$-functions to exclude the quasizero modes from  the
integral (in order to have the valley equation in the gauge sector one
of these constraints should be $\delta(A_{\mu}-A_{\mu}^v,\omega
\partial A_{\mu}^v/\partial \zeta)$ ),
make the shift $A\rightarrow A_v+A_q$, $\phi\rightarrow \phi_v+\phi_q$,
and $\psi\rightarrow\psi_v+\psi_q$ (as mentioned above fermions do not
affect the exponential such that the last transformation will not be
done explicitely), and expand the action in powers of $A_q$ and
$\phi_q$. As we shall demonstrate below, due to the valley equation in
the gauge sector, the $I\bar I$ configuration (3.3) is the approximate
valley for the gauge-Higgs model (3.2) up to order $\epsilon^2$.
Since we neglect the hard-hard and hard-soft corrections (see the
discussion in sect. 1) we can insert in eq.(3.1) just the Fourier
transform of the valley field (2.26). With exponential accuracy of
order $\epsilon^2$ the answer has the form:
\begin{equation}
A(p,k)~\sim ~ N^{-1} ~\int d\rho_1~d\rho_2 ~ dR~
A_{\mu}^{va}(p)A_{\nu}^{vb}(k)
A_{\mu}^{va}(-p)A_{\nu}^{vb}(-k)~e^{-S_v(\rho_1,\rho_2,R)}
\end{equation}
where
\begin{equation}
\left. A_{\mu}^v(p)
\right|_{p^2\rightarrow 0}= {1\over p^2} {2\pi^2\over g}
\left( \rho_1^2(\sigma_{\mu}\bar p - p_{\mu})+
\rho_2^2R(\bar \sigma_{\mu} p - p_{\mu})\bar R {e^{ipR}\over R^2} +
 O(\rho^2/\zeta)\right)
\end{equation}
\begin{eqnarray}
S_v\equiv &&  S(\phi_v,A_v)={16\pi^2\over g^2} \left(
1-6{\rho_1^2\rho_2^2\over R^4}
+12{\rho_1^2\rho_2^2\over R^6}(\rho_1^2+\rho_2^2)\right)
\nonumber \\
&&+~ \pi^2v^2(\rho_1^2+\rho_2^2)~+~2\pi^2v^2{\rho_1^2\rho_2^2\over R^2} ~+~
O({\rho^8\over g^2R^8}, {v^2 \rho^6\over R^4},\lambda v^4)
\end{eqnarray}
The first term in eq.(3.6) comes from the gauge part of the action
while the second term comes from the $\left|\nabla\phi\right|^2$
term (see ref.[10] and eq.(3.42) below). The contribution of the
$\lambda(\left|\phi\right|^2-v^2/2)^2$ term is of order $\lambda v^4\rho^4$ and
therefore exceeds our accuracy (see below). The BNV cross section is
obtained from (3.4) by analytic continuation to Minkowski energies
$E=p_0+k_0$.
\begin{equation}
\sigma_{BNV} \sim {\cal I}m \int d\rho_1 d\rho_2 dR~ \exp (ER_0-S_v
(\rho_1,\rho_2,R))
\end{equation}
In next to leading order we do not need to account for the shift of the
saddle point (1.2). Thus we can insert the
saddle values for $\rho_1$, $\rho_2$,
and $R$ into the integral (3.7) which gives the $\epsilon^2$ term in
the expansion of the
holy grail function (1.5).

Last but not least, we should prove that the configuration (3.3) is
indeed an approximate valley in $\epsilon^2$ accuracy. As discussed in
the previous section, the simplest way to prove this is by estimating
the possible additions to $S_v$ due to the linear terms $(A_{\mu}^q,
\delta S/\delta A_{\mu}^v)$,
$(\bar \phi^q,
\delta S/\delta \bar \phi^v)$, and
$(\delta S/\delta \phi^v, \phi^q)$, and
make sure that they are of higher order. After Gaussian integration of
the linear terms we have (similarly to (2.17))
\begin{equation}
S_v \rightarrow S_v-{1\over 2} \left( J_{\mu}^a\left| (\Box^{-1})_{\mu\nu}
^{ab} \right|J_{\nu}^b \right) -
{1\over 2} \left( \bar J\left|(\Box^{-1})
 \right| J \right)
\end{equation}
with
$J_{\mu}^a=\delta S/\delta A_{\mu}^a$,
$\bar J=\delta S/\delta \phi,$ and
$J=\delta S/\delta \bar \phi.$
Here $\Box^{-1}$ is the Green function of the operator $\Box\equiv
\delta^2S/\delta\bar \phi\delta \phi= -\nabla^2+4\lambda\bar \phi\phi$ and
$(\Box^{-1})_{\mu\nu}
^{ab}$ is the constraint Green function of the operator
$\Box_{\mu\nu}
^{ab}\equiv \delta^2S/\delta A_{\mu}^a\delta A_{\nu}^b= -{\cal D}^2
\delta _{\mu\nu}+2iG_{\mu\nu}+{1\over 2} g^2\bar \phi\phi$
(in the background Feynman gauge). One
of the constraints is given by the valley equation in the gauge sector
of the theory and the others could be taken as linear combinations of
the derivatives of $A_{\mu}^v$ with respect to other valley parameters
which are orthogonal to $\zeta$ (at large separations $2\zeta {d\over
d\zeta} \approx -\rho_1{d\over d\rho_1}
-\rho_2{d\over d\rho_2} $). Also, the Green functions in the background
of weakly interacting $I$ and $\bar I$ are given by the cluster
expansion (see e.g. ref [26]):
\begin{equation}
{1\over \Box} = {1\over -\partial^2+m_H^2} + \left(
{1\over \Box_I}- {1\over -\partial^2+m_H^2} \right) +
\left( {1\over \Box_{\bar I}}- {1\over -\partial^2+m_H^2} \right) +~ ...
\end{equation}
\begin{eqnarray}
\left( {1\over \Box}\right)_{\mu\nu}^{ab} &=&{\delta_{\mu\nu}\delta^{ab}
\over -\partial^2+m_W^2} + \left(
\left({1\over \Box_I}\right)
_{\mu\nu}^{ab} - {\delta_{\mu\nu}\delta^{ab}
\over -\partial^2+m_W^2} \right)
\nonumber \\
&+&
\left(
\left( {1\over \Box_{\bar I}}\right)
_{\mu\nu}^{ab} - {\delta_{\mu\nu}\delta^{ab}
\over -\partial^2+m_W^2} \right) +~ ...
\end{eqnarray}
where the dots stand for the higher terms in the expansion which are
$\sim\rho^2/R^2$.
Here $(-\partial^2+m^2)^{-1}$ are the bare propagators for the $W$ and
Higgs particles
($m_W=gv/2,~ m_H=v\sqrt{2\lambda}$) and
$(1/\Box_I)_{\mu\nu}^{ab}$, $1/ \Box_I$ are the corresponding
propagators in the field of a single instanton (similarly for $\bar
I$). Since the measure $\omega(x,\zeta)$ is approximately $1/\rho_1^2$
near the instanton (at $x^2\sim \rho_1^2$) and $1/\rho_2^2$ near the
antiinstanton (at $(x-R)^2\sim \rho_2^2$) the Green functions
$(1/\Box_I)_{\mu\nu}^{ab}$, and
$(1/\Box_{\bar I})_{\mu\nu}^{ab}$ could be taken as the BCCL
propagators [27] with the restrictions proportional to the pure zero
modes. (We shall not, however, need the explicit form of these BCCL
propagators.)

Now let us estimate the additional term in (3.8) at the saddle point
values for $\rho$ and $R$ (1.2). The first functional derivatives are
\begin{equation}
J_{\alpha}^a \equiv {\delta S\over \delta A_{\alpha}^a} =
-{\cal D}_{\mu} G_{\mu\alpha}^a ~+~ ig\bar \phi
\{ t^a, \nabla_{\alpha}\}\phi
\end{equation}
\begin{equation}
{\cal D}_{\mu}G_{\mu\alpha}^v ~=~ {24i\over g}~ {\rho_1^2\rho_2^2\over
R^2}~ { x\bar \sigma_{\alpha}(x-R)\bar R - (trace)\over
x^4(x-R)^4\Pi_1^2\Pi_2^2} ~+~...
\end{equation}
\begin{eqnarray}
t^a \cdot ig\bar \phi_v(t^a \nabla_{\alpha}^v
+\stackrel{\leftarrow}{\nabla}_{\alpha}^vt^a)\phi_v=
-{igv^2\over 4}~\left( {\rho_1^2(\sigma_{\mu}\bar x-x_{\mu})\over
x^4\Pi_1^2\Pi_2} \right.
\nonumber \\
\left. +{\rho_2^2R(\bar \sigma_{\mu}(x-R)-(x-R)_{\mu})\bar R \over
R^2(x-R)^4 \Pi_1 \Pi_2^2}\right) ~+...
\end{eqnarray}
and
\begin{equation}
J\equiv {\delta S\over \delta \bar \phi}=-\nabla^2\phi+2\lambda(\bar
\phi \phi-v^2/2)\phi
{}~~~,~~~
\bar
 J\equiv {\delta S\over \delta  \phi}=-\bar
\phi\stackrel{\leftarrow}{\nabla}^2+2\lambda(\bar
\phi \phi-v^2/2)\bar \phi
\end{equation}
\begin{eqnarray}
(\nabla^2\phi)_v= {4ve\over \sqrt{2\Pi_1\Pi_2}}~{\rho_1^2
\rho_2^2 \over R^2x^4(x-R)^4 \Pi_1\Pi_2}~\left\{ 2~x\cdot R~
(x-R)\cdot R~
 + 2(\rho_1^2+\rho_2^2)
\right. \nonumber \\
\left.
(x^2-(x\cdot R)^2/R^2)-
(x \bar R -~x \cdot R)(R^2+x^2+(x-R)^2)
\right\} + ...~~~~~~
\end{eqnarray}
\begin{equation}
2\lambda(\bar \phi_v\phi_v-v^2/2)\phi_v ~=~ -{\lambda v^3 e
\over \sqrt{2\Pi_1\Pi_2}}~ \left( {\rho_1^2\over x^2}+
{\rho_2^2\over (x-R)^2} \right) ~+~ ...
\end{equation}
and similarly for $\bar J$ (as usual dots stand for higher orders in
$\rho^2/R^2$).
The double integrals in (3.8) are of the type $\int dxdy~
J(x)G(x,y)J(y)$ which encompass three characteristic regions of
integration: (i) $x,y\sim \rho_1$ (or $x-R, y-R\sim \rho_2$), (ii)
$x,y\sim R$, and
(iii) $x,y\sim 1/m\gg R$. We first consider the contribution to
eq.(3.8) comming from the first region. Here
\begin{equation}
J_{\mu}~=~ {i\over g}~
 { \sigma_{\mu}\bar x -x_{\mu}\over
x^4\Pi_1^2} ~\left( {24\over \zeta^2}-{g^2v^2\over 2}\rho_1^2 \right)
\end{equation}
\begin{equation}
J~=~ {8\over \zeta^2}~
 { x\bar R\over
x^4\Pi_1} ~{ve\over \sqrt{2\Pi_1}}
\end{equation}
and the Green functions (3.9) and (3.10)are the BCCL propagators in
the instanton background.From dimensional considerations it follows
\begin{equation}
(J_{\mu}|({1\over \Box})_{\mu\nu}|J_{\nu}) ~\sim ~ {1\over g^2}
(24\zeta^{-2}-2m_w^2\rho_1^2) ~\sim~ {1\over g^2} \epsilon ^{8/3}
\end{equation}
\begin{equation}
(\bar J|({1\over \Box})|J) ~\sim ~
v^2R^2/\zeta^4 ~\sim~ {1\over g^2} \epsilon ^{10/3}
\end{equation}
In fact the expression (3.19) is also $\sim \epsilon^{10/3}g^{-2}$
since $J_{\mu}$ is proportional to $\rho_1\partial A_{\mu}/d\rho_1$ which
is one of the instanton zero modes while $(1/\Box)_{\mu\nu}$ is the BCCL
propagator orthogonal to the zero modes (and the dilatation mode
$\partial A/\partial \rho$ in particular). Let us consider next the
contribution from region (iii) where the Green functions (3.9) and
(3.10) are bare propagators.
\begin{equation}
J_{\mu}~=~ -{i\over g}~{m_W^2\over x^4}~
[ \rho_1^2 ( \sigma_{\mu}\bar x -x_{\mu}) + \rho_2^2
R( \bar \sigma_{\mu} x -x_{\mu})\bar R] + ...
\end{equation}
\begin{equation}
J~=~ -{m_H^2\over 2}~
 {\rho_1^2+\rho_2^2\over x^2}~
{ve\over \sqrt{2}}~+...
\end{equation}
Again dimensional considerations give
\begin{equation}
(J_{\mu}^a |{1\over -\partial^2+m_W^2} |J_{\mu}^a) ~\sim ~
{1\over g^2} m_w^4\rho^4 ~\sim~ {1\over g^2}\epsilon^{8/3}
\end{equation}
\begin{equation}
(\bar{J}|{1\over -\partial^2+m_H^2} |J) ~\sim ~
m_H^4\rho^4v^2\cdot {1\over m_H^2} ~\sim~ {1\over g^2}\epsilon^{8/3}
\end{equation}
Similarly, it is easy to verify that the contributions from the region
$x,y\sim R$, where the propagators (3.9) and (3.10) are bare and
$J_{\mu}$ and $J$ are given by (3.11)-(3.16), is also
$ {1\over g^2}\epsilon^{8/3}$.
This proves that (3.3) is a valley solution up to order $\epsilon^2$
but n~o~t in higher orders as the linear terms give then additional
contributions.

However, one can modify the configuration (3.3) such that it becomes a
valley at least up to the order $\epsilon^{8/3}\ln \epsilon$ (the
deviations are of order $\epsilon^{8/3}$). Since all the contributions
to eq.(3.8) $\sim \epsilon^{8/3}$  come from the terms (3.23) and
(3.24) and are proportional to $m_W^2$ and $m_H^2$ the valley has to be
modified such that it accounts for the gauge boson and Higgs masses.
This implies that the valley configuration should decrease
exponentially at large $x^2\gg R^2$ with the Higgs and boson masses
setting the scale. In other words, the improved valley has to be
constructed from constraint instantons (see ref.[28]) rather than from
ordinary ones. For the accuracy we aim at it is sufficient to modify
the valley (3.3) in the following manner
\begin{equation}
A_{\mu}^v= A_{\mu}^I + A_{\mu}^{\bar I} + B_{\mu} ~~,~~
\phi_v=\phi_1\phi_2{ve\over \sqrt{2}}~~,~~
\bar \phi_v=\phi_1\phi_2{v\bar e\over \sqrt{2}}~~,~~
\end{equation}
where $I$ and $\bar I$ are now constraint instantons [28]:
\begin{equation}
A_{\mu}^I=-{i\over g} \rho_1^2{\sigma_{\mu}\bar x-x_{\mu}\over
\Pi_1}~ \left\{ {\Theta(R^2-x^2)\over x^4}-\Theta(x^2-R^2)G_W'(x^2)
\right\}
\end{equation}
\begin{eqnarray}
A_{\mu}^{\bar I}&=&-{i\over g} \rho_2^2{R(\bar
\sigma_{\mu}(x-R)-(x-R)_{\mu})\bar R\over R^2
\Pi_2}~ \left\{ {\Theta(R^2-(x-R)^2)\over (x-R)^4}- \right.
\nonumber \\
&&~~~~~~~~~~~~~~~\Theta((x-R)^2-R^2)G_W'((x-R)^2)
\Big{\}}
\end{eqnarray}
\begin{equation}
\phi_1={\Theta(R^2-x^2)\over \sqrt{\Pi_1}}
+\Theta(x^2-R^2)\sqrt{1-\rho_1^2\Pi_1^{-1}G_H(x^2)}
\end{equation}
\begin{equation}
\phi_1={\Theta(R^2-(x-R)^2)\over \sqrt{\Pi_2}}
+\Theta((x-R)^2-R^2)\sqrt{1-\rho_2^2\Pi_2^{-1}G_H((x-R)^2)}
\end{equation}
Here $G_W$ and $G_H$ are the bare propagators with masses $m_W$ and
$m_H$ respectively:
\begin{equation}
G_W(x^2)= \int{dp\over 4\pi^2}~ {e^{-ipx}\over m_W^2+p^2}~~~,~~~
G_H(x^2)= \int{dp\over 4\pi^2}~ {e^{-ipx}\over m_H^2+p^2}
\end{equation}
and $G_W'(x^2)$ is the derivative of $G_W(x^2)$ with respect to $x^2$.
The field $B_{\mu}$ should also be modified to become exponentially
decreasing but since $B_{\mu}$ itself is small ($\sim \rho^2/R^2$) its
mass dependence is not essential at our accuracy. For vanishing masses
the new valley (3.25) obviously reduces to the configuration (3.3). For
large $x^2$, on the other hand, it corresponds to the emission of
massive particles. Let us demonstrate now that for the improved
configuration all additional contributions to $S_v$ (due to linear
terms) are of higher order than $\epsilon^{8/3}\ln \epsilon$.

Logarithmic contributions of the order $\epsilon^{8/3}\ln \epsilon$ can
come from two regions of integration in $x$ and $y$ :
(1) $R^2\gg x^2,y^2\gg \rho_1^2$( or $R^2\gg (x-R)^2,(y-R)^2\gg \rho_2^2$) and
(2) $m^{-2}\gg x^2,y^2\gg R^2$
(as usual we assume that $m_H$ is of the same order of magnitude as
$m_W$). In the first region the valley configuration coincides with the
massless valley (3.3), $J_{\mu}$ and $J$ are given by eqs. (3.17) and
(3.18) and the Green functions (3.9) and (3.10) are the BCCL
propagators in the $I(\bar I)$ background. Strictly speaking in region
(1) we should use the large $x^2$ asymptotics of these expressions.
Since the BCCL propagator contains explicitely the logarithmic
term $\sim \Phi_0(x)\Phi_0(y)\ln (R^2/\rho^2)$ it looks as if
the region $x^2,y^2\sim
\rho^2$ were also essential.
However, the contribution to $(J_{\mu}(\Box^{-1})_{\mu\nu}
J_{\nu})$ coming from this region vanishes because $J_{\mu}$ is proportional
to $\partial A_{\mu}^I/\partial \rho$ which is one of the constraints
of the BCCL propagator.
For the same reason no $\epsilon^{8/3}\cdot const$ term arises from
the region $x^2,y^2\sim \rho^2$. It could, however, result from the
region $x^2,y^2\sim R^2$. The term $(\bar J \Box^{-1} J)$ does not
contain any such contribution $\sim \epsilon^{8/3}$ (see eq.(3.20))
and is therefore unimportant.

In the second region ($m_H^{-2}\gg x^2,y^2 \gg R^2$) all fields
decrease exponentially, so
\begin{equation}
J_{\mu}^v=-(\partial^2\delta_{\mu\nu}-\partial_{\mu}\partial_{\nu})
A_{\nu}^v+t^a~{g^2v^2 \over 2}\bar e~ \{t^a,A_{\mu} \}e=
-(m_W^2-g^2v^2/4)A_{\mu}^v
\end{equation}
\begin{eqnarray}
J=&& -\partial^2\phi_v+2\lambda \left( \bar \phi_v\phi_v-
{v^2\over 2}\right) \phi_v=
-(m_H^2-2\lambda v^2)(\phi_v-{ve\over \sqrt{2}})
\nonumber \\
\bar J=&&
-(m_H^2-2\lambda v^2)(\bar \phi_v-{v \bar e\over \sqrt{2}})
\end{eqnarray}
where $A_{\mu}^v$ and $\phi_v$ are given by the asymptotics of
eq. (3.25) at large $x^2$.  Since in this region the Green functions (3.9)
and (3.10) reduce to the bare propagators we obtain
\begin{eqnarray}
\left( J_{\mu}^{va}~| \right.&
\left. {1\over -\partial^2+m_W^2}~|~J_{\mu}^{va} \right)=
-\left( m_W^2-{g^2v^2\over4} \right)^2~{12(\rho_1^4+\rho_2^4)\over
g^2}~ \int {dp~p^2\over (m_W^2+p^2)^3} \nonumber \\
&\simeq
-\left( m_W^2-{g^2v^2\over4} \right)^2~{12\pi^2(\rho_1^4+\rho_2^4)\over
g^2}~ \ln  {1\over m_W^2R^2}
\end{eqnarray}
\begin{eqnarray}
\left(\bar J~| \right. &
\left. {1\over -\partial^2+m_H^2}~|~J \right)=
( m_H^2-2\lambda v^2)^2~{v^2(\rho_1^4+\rho_2^4)\over
8}~ \int {dp~\over (m_H^2+p^2)^3} \nonumber \\
&=
( m_H^2-2\lambda v^2)^2~{\pi^2v^2(\rho_1^4+\rho_2^4)\over
8m_H^2}\sim {m_W^2m_H^2\rho^4\over g^2}
\end{eqnarray}
We see now that for the properly chosen mass $m_W^2=g^2v^2/4$
the logarithmic contribution in eq.(3.33)
vanishes. As eq.(3.34) does not contain any such term from the beginning
we conclude that the configuration (3.25) is a proper valley up to
$\epsilon^{8/3}\cdot const$ terms.

Finally, in order to find the $\epsilon^{8/3}\ln \epsilon$ contribution
to the holy grail function (1.4) we have to calculate the action of the
valley configuration (3.25) at the saddle point values of
$\rho_1,\rho_2$ and $R$. (If we were interested in the
$\epsilon^{8/3}\cdot const$ terms we would also have to account for the
shift of the saddle point due to the next-to-leading terms $\sim
g^{-2}\rho^6R^{-6}$ and $v^2\rho^4R^{-2}$ (see eq.(3.6)). It is
convenient to expand the action of the valley configuration (3.25) as a
power series in $m_W^2$ and $m_H^2$ since we need only the first few
terms of this expansion (see below). Let us start with the gauge part
of the action $S_g=\int dx {1\over 4} G_{\mu\nu}^aG_{\mu\nu}^a$.
\begin{eqnarray}
S_g^v&=&\left. S_g^v \right|_{m_W^2=0}-m_W^2
\left. \left( {\partial A_{\alpha}^
{va}\over \partial m_W^2},{\cal D}_{\mu}G_{\mu\alpha}^{va} \right)
\right|_{m_W^2=0}
\nonumber \\
&+&{m_W^4\over 2}\left\{ \left( {\partial^2 A_{\alpha}^
{va}\over \partial (m_W^2)^2},-{\cal D}_{\mu}G_{\mu\alpha}^{va}\right)
\right.
\nonumber \\
&+&\left. \left. \left( {\partial A_{\mu}^
{va}\over \partial m_W^2},(-{\cal D}^2\delta_{\mu\nu} +2iG_{\mu\nu}
\right) ^{ab}
{\partial A_{\nu}^
{vb}\over \partial m_W^2} \right\} \right|_{m_W^2=0} + ...
\end{eqnarray}
The first term is the action of the massless valley up to order ${
\rho^8\over R^8}\ln {R^2\over \rho^2}$.
\begin{equation}
\left. S_g^v \right|_{m_W^2=0}= {16\pi^2\over g^2} \left( 1-6
{\rho_1^2\rho_2^2\over R^4}+
12{\rho_1^2+\rho_2^2\over R^6}
\rho_1^2\rho_2^2+
36{\rho_1^4\rho_2^4\over R^8}
\ln {R^2\over \rho_1\rho_2}+...\right)
\end{equation}
where the last term gives in the saddle point
a $\epsilon^{8/3}\ln \epsilon$ contribution
to the holy grail function.

We shall discuss next the other terms in (3.35). All the
higher terms in the $m_W^2$ expansion contribute only
in higher orders. This is easy to see, as the dimension of  $m_W^n$
is allways balanced by a factor proportional to $\rho_{1,2}^n \sim
\epsilon^{2n/3}$.
Since
\begin{eqnarray}
&&\left. {\partial A_{\nu}^
v\over \partial m_W^2}\right|_{m_W^2=0} =
{i\rho_1^2\over 4g}{\sigma_{\mu}\bar x-x_{\mu}\over x^2\Pi_1}
\Theta(x^2-R^2) \nonumber \\
&&+
{i\rho_2^2\over 4g}{R(\bar \sigma_{\mu}
(x-R)-(x-R)_{\mu})\bar R\over R^2(x-R)^2\Pi_2}
\Theta((x-R)^2-R^2)
\end{eqnarray}
we obtain
\begin{eqnarray}
m_W^2\left( {\partial A_{\alpha}^
{va}\over \partial m_W^2} ,{\cal D}_{\mu}G_{\mu\alpha}^{va} \right)
&=& -{72m_W^2\over g^2} {\rho_1^2\rho_2^2\over R^2}~ \int dx~
\left\{ \rho_1^2~ (x-R)\cdot R~ \Theta(x^2-R^2)
\right. \nonumber \\
&+&\left. \rho_2^2~ x\cdot R~ \Theta((x-R)^2-R^2)\right\} x^{-4}(x-R)^{-4}
\nonumber \\
&\sim & {m_W^2\rho_1^2\rho_2^2\over g^2R^4}(\rho_1^2+\rho_2^2)
\sim \epsilon^{8/3}
\end{eqnarray}
Similarly we can show that the first term of the $m_W^4$
contribution in eq.(3.35) is of higher order
\begin{equation}
-m_W^4\left( {\partial^2 A_{\alpha}^
{va}\over \partial (m_W^2)^2},{\cal D}_{\mu}G_{\mu\alpha}^{va}\right)
{}~\sim~ m_W^4{\rho_1^2\rho_2^2\over R^2}(\rho_1^2+\rho_2^2) ~\sim~
\epsilon^{10/3
}
\end{equation}
and we are left with the second term of this contribution. Since
all fields vanish exponentially at
large $x^2$ we can replace the covariant derivatives
by the ordinary ones.
\begin{eqnarray}
{m_W^4\over 2}&& \left( {\partial A^{va}\over \partial m_W^2},
(-\partial^2)
{\partial A^{va}\over \partial m_W^2} \right) =
{m_W^4\over 2g^2} \int dx~\left\{ {3\rho_1^4\over x^4}\Theta(x^2-R^2)
\right. \nonumber \\
&+& {3\rho_2^4\over (x-R)^4}\Theta((x-R)^2-R^2)
-{2\rho_1^2 \rho_2^2\over R^2x^2(x-R)^2}~
\left({1\over x^2}+{1\over (x-R)^2}\right) ~
\nonumber \\
&&
(4~x\cdot R~(x-R)\cdot R-
 R^2~x\cdot (x-R))\Theta(x^2-R^2)
\Theta((x-R)^2-R^2)\Big{\}}
\nonumber \\
&=& {3\pi^2m_W^4\over 2g^2}
(\rho_1^4+\rho_2^4)\ln{1\over m_W^2R^2}
\sim \epsilon^{8/3}\ln \epsilon
\end{eqnarray}
(The upper limit of the logarithmic integral is $1/m_W^2$ and the lower
one is $R^2$ due to the $\Theta$-function.)
This term contributes to the holy grail function in the order we are
interested in.

Next we consider the gauge-Higgs part of the action $S_{gH}=\int dx
|\nabla\Phi|^2$ and expand it in powers of $m_W^2$ and $m_H^2$.
\begin{eqnarray}
S_{gH}=&\left. S_{gH}\right|_{m_H,m_W=0}+
\left. m_W^2ig\left( {\partial A_{\mu}^a\over \partial m_W^2},\bar \phi
\{t^a\nabla_{\mu}\}\phi\right)\right|_{m_W,m_H=0}^v\\
&\left.- m_H^2\left\{ \left( {\partial \bar \phi\over \partial m_H^2},
\nabla^2\phi \right) + \left( \bar \phi \stackrel{\leftarrow}
{\nabla}^2,
{\partial \phi\over \partial m_H^2}\right) \right\}
\right|_{m_W,m_H=0}^v+...~~~
\nonumber
\end{eqnarray}
The best way to calculate the first term is to use the explicit expression
(3.15)
for $\nabla^2\Phi$ for the massless valley. We have
\begin{eqnarray}
S_{gH}&=&\pi^2v^2(\rho_1^2+\rho_2^2)-4v^2{\rho_1^2\rho_2^2\over R^2}~
 \int dx~ { x\cdot R~ (x-R)\cdot R\over x^4(x-R)^4\Pi_1^2\Pi_2^2}
 \\
&=&\pi^2v^2(\rho_1^2+\rho_2^2) + 2\pi^2v^2{\rho_1^2\rho_2^2\over R^2}-
6\pi^2v^2{\rho_1^2\rho_2^2\over R^4}~ \left( \rho_1^2\ln
{R^2\over \rho_1^2}+
\rho_2^2 \ln {R^2\over \rho_2^2}\right)\nonumber
\end{eqnarray}
{}From the three terms in the braces on the r.h.s. of eq.(3.15) only the
first contributes. The second is of higher order in $\epsilon$ and the
third one vanishes after integration over x. The first term on the
r.h.s. of eq.(3.42) is the surface term due to partial integration :
$\int dx~\left|\nabla\phi_v\right|^2~=\pi^2v^2(\rho_1^2+\rho_2^2) - \int dx
 ~\bar
\phi_v \nabla^2 \phi_v$.

Next, let us address the $m_W^2$ term on the r.h.s. of eq.(3.41). With
(3.13) and (3.37) we easily obtain the same integral as in eq.(3.40)
(but with different coefficient).
\begin{eqnarray}
&&\left. m_W^2ig\left( {\partial A_{\mu}^a\over \partial m_W^2},\bar \phi
\{ t^a\nabla_{\mu}\}\phi\right) \right|_{m_W,m_H=0}^v
=-{v^2m_W^2\over 4}~ \int dx ~\left\{ {3\rho_1^4\over x^4} \Theta(x^2-R^2)+
\right. \nonumber \\
&&\left. {3\rho_2^4\over (x-R)^4} \Theta((x-R)^2-R^2)+ ...\right\}
= -{3\pi^2\over 4}m_W^2v^2(\rho_1^4+\rho_2^4)\ln{1\over m_W^2R^2}
\end{eqnarray}
The $m_H^2$ term in eq.(3.41) does not contribute at our accuracy.
By inserting the explicite form of $\nabla^2\Phi_v$ (3.15) one can
convince oneself that it only contributes in the order
$\rho_{1,2}^6R^{-2}\sim \epsilon^{10/3}$.

Finally we come to the last part of the action, namely the Higgs
self-interaction.
\begin{eqnarray}
S_H&=& \int dx \lambda(\bar \phi_v\phi_v-v^2/2)^2= {\lambda v^4\over 4}
\int dx  \left\{ {\rho_1^4\over \Pi_1^2} [\Theta(R^2-x^2)x^{-4}
\right. \nonumber \\
&&~~~~~~~~~~~~~~~+\Theta(x^2-R^2)G^2_H(x^2)]
\nonumber \\
&+&  {\rho_2^4\over \Pi_2^2} [\Theta(R^2-(x-R)^2)(x-R)^{-4}
+\Theta((x-R)^2-R^2)G^2_H((x-R)^2)]+ \nonumber \\
&+&  2\rho_1^2\rho_2^2 G_H(x^2)G_H((x-R)^2)\Theta(x^2-R^2)
\Theta((x-R)^2-R^2) \Big{\}} \nonumber \\
&=& {\lambda v^4\pi^2\over 4} \left( \rho_1^4\ln {1\over
m_H^2\rho_1^2}
+\rho_2^4\ln {1\over m_H^2\rho_2^2}
+\rho_1^2\rho_2^2\ln {1\over m_H^2R^2} \right)
\end{eqnarray}
Combining now all our results, namely (3.6), (3.36), (3.40), (3.42),
(3.43), and (3.44) we obtain the final answer for the action
of our massive valley (3.25).
\begin{eqnarray}
S_v&=& {16\pi^2\over g^2} \left\{ 1
-6{\rho_1^2\rho_2^2\over R^4}
+12{\rho_1^2\rho_2^2\over R^6}(\rho_1^2+\rho_2^2)
+36{\rho_1^4\rho_2^4\over R^8}\ln {R^2\over \rho_1\rho_2}
\right. \nonumber \\
&&~~~~~~~~~~~~~~~\left.
+{3 m_W^4\over 32}(\rho_1^4+\rho_2^4) \ln {1\over m_W^2R^2}\right\}
\nonumber \\
&&+ \pi^2 v^2 \left\{
\rho_1^2+\rho_2^2
+2{\rho_1^2\rho_2^2\over R^2}
-6{\rho_1^2\rho_2^2\over R^4}\left( \rho_1^2\ln {R^2\over \rho_1^2}
+\rho_2^2\ln {R^2\over \rho_2^2} \right)
\right. \nonumber \\
&&~~~~~~~~~~~~~~~\left.
-{3 m_W^2\over 4}(\rho_1^4+\rho_2^4) \ln {1\over m_W^2R^2}\right\}
\nonumber \\
&&+ {\lambda \pi^2 v^4\over 4} \left\{
\rho_1^4\ln {1\over m_H^2\rho_1^2}
+\rho_2^4\ln {1\over m_H^2\rho_2^2}
\right. \nonumber \\
&&~~~~~~~~~~~~~~~\left.
+2\rho_1^2\rho_2^2\ln {1\over m_H^2R^2}\right\}  ~+...
\end{eqnarray}
Substituting now the saddle point vallues (1.2) for $\rho_1$, $\rho_2$,
and $R$ we obtain the expansion (1.5) of the holy grail function.
(Recall that we assume $m_H$ to be of
order $m_W$ so we do not distinguish between $\ln m_H^2$ and $\ln
m_W^2$. The mass difference enters only at order
$\epsilon^{8/3}\cdot$
const.)

\vskip 1 cm
\setcounter{chapter}{4}
\setcounter{equation}{0}
\LARGE
4. Effective Lagrangean for instanton-induced interactions.
\normalsize
\vskip 0.6 cm

As we discussed in the Introduction, the basic assumption of the effective
Lagrangean approach is  that the instanton-induced processes are
given by the ordinary perturbative diagrams and that additional
multiparticle vertices originate from the effective Lagrangean
\begin{equation}
L_{\rm eff}=L_{\rm eff}^I+
L_{\rm eff}^{\bar I}~~~~,~~~~
L_{\rm eff}^{I(\bar I)}(z)=\int {d\rho\over \rho^5}d(\rho)
L_{\psi}^{I(\bar I)}(\rho,u) \exp\left(
-L^{I(\bar I)} (z)\right)
\end{equation}
where
$d(\rho)\sim exp(-{8\pi^2\over g_W^2})$ is the usual instanton
density[1].Here $L_{\psi}^{I(\bar I)}$ is
the t'Hooft effective Lagrangean for
fermions in the (anti)instanton field [1]
\begin{eqnarray}
L_{\psi}^I& = (4\pi^2\rho_I^3)^6~ \prod_{k=1}^6~ (\psi^{k+6},u_0\bar{u
}_I \epsilon)(\epsilon {u}_I \bar u_0,\psi^k)~+...
\nonumber \\
L_{\psi}^{\bar I}& = (4\pi^2\rho_{\bar{I}}^3)^6~ \prod_{k=1}^6~ (\bar\psi_k,u
_{\bar I}\epsilon)(\epsilon{u}_{\bar I}, \bar \psi_{k+6})~+...
\end{eqnarray}
and $L^{I(\bar I)}$ is the instanton-induced effective Lagrangian for
 bosons[11,12]
\begin{eqnarray}
L^I& = -{2\pi^2i\over g} \rho_I^2~ {\cal T}r\{ u_0\bar{u_I}
\sigma_{\alpha} \bar \sigma_{\beta} u_I
\bar{u}_0 G_{\alpha \beta}\} ~+2\pi^2\rho_I^2\bar \phi\phi~ +...
\nonumber \\
L^{\bar I}& = -{2\pi^2i\over g} \rho_{\bar I}^2~ {\cal T}r\{ u_{\bar I}I \bar
\sigma_{\alpha}  \sigma_{\beta}
\bar u_{\bar I}G_{\alpha \beta}\} ~+2\pi^2\rho_{\bar I}^2\bar \phi\phi~ +...
\end{eqnarray}
($u_0$ is an arbitrary unit vector which drops from the final results for
physical amplitudes). As usual $(\epsilon u)$ denotes
$\epsilon_{\alpha\beta}u^{\beta\gamma}$ etc. . The
ellipsis stand for operators of higher dimensions, multiplied by
additional powers of $\rho$. (Some of the next to leading terms $\sim
\rho^4$ are given below, see eq.(4.26).) Eqs. (4.2) and (4.3) are
infinite series of local operators with increasing dimension,
multiplied by growing powers of $\rho$.

The effective Lagrangean (4.1) added to the ordinary one reproduces
the instanton-induced effects. More precisely, if we start from the
Lagrangean
\begin{equation}
L={1\over 4} G^a_{\mu\nu}G^a_{\mu\nu}+ \left| \nabla \phi\right|^2 +
\lambda(\left|\phi\right|^2-v^2/2)^2+
L_{\rm eff}^I+
L_{\rm eff}^{\bar I}
\end{equation}
and expand up to the m-th power in
$L_{\rm eff}^I$ and n-th power in
$L_{\rm eff}^{\bar I}$, sum up the corresponding perturbative
diagrams with m $I$-type vertices and n $\bar I$-type vertices (each
of them couples 12 fermions and an arbitrary number of W's and H's),
we should reproduce the answer for the original amplitude calculated
with a background of m instantons and n antiinstantons.
To illustrate this equivalence let us consider the simplest example,
namely just one instanton and let us neglect for the moment the W and
H masses. One instanton effects are described by the first term in
the power expansion of $\int dz L_{\rm eff}^I(z)$. Thus $\langle
A_{\mu}(x)e^{ -L^I(z)}\rangle$ should reproduce an instanton field
with size $\rho$, center $z$ and orientation matrix $u \bar u_0$ (in
the singular gauge):
\begin{eqnarray}
\left\langle A_{\mu}(x)\exp\left( {2\pi i\over g}
\rho^2 {\cal T}r\{u_0\bar u
\sigma_{\alpha}\bar\sigma_{\beta}u\bar u_0 G_{\alpha\beta}(z)
\}\right) \right\rangle =\nonumber \\
-{i\over g} u_0\bar u(\sigma_{\mu}\bar\Delta-\Delta_{\mu})u\bar u_0
{\rho^2\over \Delta^2(\Delta^2+\rho^2)}~~~,~~~\Delta=x-z
\end{eqnarray}
Here by $\langle {\cal O}\rangle$ we denote averaging ${\cal O}$ with
exp(-$S$), $S$ being the ordinary action (3.1). Let us verify
(4.5) by expanding its l.h.s. in powers of $\rho^2G^a_{\mu\nu}$.
The first term of the expansion is simply (see Fig.7a)
\begin{equation}
\overline{A_{\mu}(x)G}^a_{\alpha\beta}(z) {2\pi i\over g} \rho^2
{\cal T}r\{
u_0\bar u\sigma_{\alpha}\bar\sigma_{\beta}u\bar u_0 t^a
\}=
-{i\over g} \rho^2 u_0\bar
u(\sigma_{\mu}\bar\Delta-\Delta_{\mu})u\bar u_0
\Delta^{-4}
\end{equation}
which gives the asymptotics of the instanton field at $\Delta^2 \gg
\rho^2$. The second term in the expansion of the l.h.s. of eq.(4.5)
corresponds to the diagram shown in Fig.7b. The calculation of this
diagram gives the second term in the expansion of the instanton field
in powers of $\rho^2/\Delta^2$, namely ${i\over g} \rho^4 u_0 \bar u
(\sigma_{\mu}\bar \Delta-\Delta_{\mu})u\bar u_0 \Delta^{-6}$. The
third term corresponds to the two diagrams in Fig7c. It can be
demonstrated that the logarithmic contributions $\sim (\ln
\Delta)^2\Delta^{-8}\rho^6$ from these diagrams cancel and the result
coincides with the third term in the expansion of the instanton field
(4.5) in powers of $\rho^2/\Delta^2$. In general
a series of diagrams of the type shown in Fig.7 describes the
perturbative solution of the classical equation ${\cal
D}_{\mu}G_{\mu\nu}=0$ using the asymptotics (4.6) as first iteration.
Since every extra factor of $g$ is compensated by an extra factor
$\rho^2/g$ coming from the emission of an additional gauge boson by
the instanton this perturbative expansion is in powers of $\rho^2$
rather than $g$.

The situation for the scalar instanton component is quite similar -
it is reproduced by $\langle \phi(x) e^{- L_I(z}\rangle$ as a power
series in $\rho^2/\Delta^2$ corresponding to the diagrams shown in
Fig. 8. Note that after the shift $\phi \rightarrow \phi +
(v/\sqrt{2}) e$ we have
\begin{eqnarray}
L^I =& -{2\pi^2i\over g} \rho_I^2~ {\cal T}r\{ u_0\bar u_I
\sigma_{\alpha} \bar \sigma_{\beta} u_I
\bar u_0 G_{\alpha \beta}\} ~+2\pi^2\rho_I^2\bar \phi\phi~ +
\pi^2v^2\rho_I^2
\nonumber \\
&+2\pi^2\rho_I^2 {v\over \sqrt{2}}(\bar \phi e+\bar e
\phi)+...
\end{eqnarray}

and similarly for $\bar I$. The asymptotical scalar field comes from
the diagram in Fig. 8b (the trivial term $(v/\sqrt{2})$ can be
depicted as in Fig. 8a).
\begin{equation}
\phi={v\over \sqrt{2}} e - 2\pi^2\rho^2 \overline{\phi(x)\bar
\phi}(z) e {v\over \sqrt{2}} =
{v\over \sqrt{2}}e\left( 1-{\rho^2\over 2\Delta^2} \right)
\end{equation}
The further diagrams of the series in Fig.8 give subsequent terms in
the expansion of the scalar component of the instanton field in
powers of $\rho^2/\Delta^2$
\begin{equation}
\langle \phi(x) e^{- L_I(z)}\rangle = {v\over \sqrt{2}}
{e\over \sqrt{1+\rho^2\Delta^{-2}}} = {v\over \sqrt{2}}
e \left( 1-{\rho^2\over 2\Delta^2} +{3\rho^4\over 8
\Delta^4}+...\right)
\end{equation}
This series corresponds to the perturbative solution of the equation
$\nabla^2\phi=0$ starting from (4.8) as the first iteration,
expressing the field $A_{\mu}^I$ as power series in $\rho^2/x^2$.

It should be mentioned that for massive particles (e.g. if the
massive propagator is inserted in Fig. 6-8) the classical fields
$\langle A_{\mu}(x) e^{- L_I(z)} \rangle$ and
$\langle \phi(x) e^{- L_I(z)} \rangle$ become the exponentially
decreasing configurations of the constraint-instanton type [28].
The expansions (4.5) and (4.9) are then only valid for $x^2\ll m^2$
while at large $x$ every term in the expansion is multiplied by a
function of $m^2x^2$ determined by the corresponding diagram. For
example, the first nontrivial terms $\sim \rho^2$ are proportional to
the bare massive propagators with $m_W$ and $m_H$, and at this order
the classical fields
$\langle A_{\mu}(x) e^{- L_I(z)} \rangle$ and
$\langle \phi(x) e^{- L_I(z)} \rangle$ coincide with our valley
configurations (3.26) and (3.28) (at $x^2> R^2$).

It is also instructive to analyse how the Green functions in the
instanton background are obtained within the effective Lagrangean
approach. For simplicity we shall consider the scalar propagator in
the background of a single instanton with center $z=0$ and
orientation matrix $u \bar u_0=1$. The explicite form of this
propagator is [28] (here we again neglect $m_W$ and $m_H$):
\begin{eqnarray}
G(x,y)&& = {1+{\rho^2 x\bar y\over x^2 y^2}\over 4\pi^2 (x-y)^2
\sqrt{\Pi_x\Pi_y}} = {1\over 4\pi^2(x-y)^2} + {\rho^2(2x\bar y - x^2
-y^2)\over 8\pi^2x^2y^2(x-y)^2} \nonumber \\
&&+{\rho^4\over 32\pi^2} \left\{ {1\over (x-y)^2} \left[ \left(
{1\over x^2}-{1\over y^2}\right)^2 -
{4(x\bar y-x\cdot y)\over x^2y^2}  \left(
{1\over x^2}-{1\over y^2}\right)\right]+\right. \nonumber \\
&&\left. +{2\over x^2y^2}\left(
{1\over x^2}-{1\over y^2}\right)\right\} +...
\end{eqnarray}
where $\Pi_x=1+\rho^2x^{-2}$,
$\Pi_y=1+\rho^2y^{-2}$ and the ellipsis stands for higher order terms
in $\rho^2$. On the other hand, using the effective Lagrangean this
propagator can be represented as
\begin{equation}
\langle \phi(x) \bar \phi(y) \exp -L_I(0)\rangle
\end{equation}
or, in detail, by the sum of diagrams in Fig.9 corresponding to an
expansion of (4.11) in powers of $L^I$ (i.e. in powers of $\rho^2$,
see eq.(4.7)). Apart from the bare propagator (shown in Fig. 9a) the
first non-trivial term $\sim \rho^2$ comes from the two diagrams in
Fig. 9b. The first diagram is generated by the $\rho^2G$ term in the
effective Lagrangean (4.7). The calculation yields
\begin{equation}
{\rho^2 (x\bar y - x\cdot y)\over 4\pi^2 x^2 y^2 (x-y)^2}
\end{equation}
(To verify this result it is easiest to differentiate (4.12) to amputate
one leg of the first diagram in Fig.9b, i.e. to calculate $\partial^2/\partial
x_{\alpha} \partial x_{\alpha}$ of eq.(4.12)). The second contact-type
diagram is obtained when we take the last term in the effective
Lagrangean (4.7) $\sim \rho^2\bar \phi \phi$. The result is
\begin{equation}
-{\rho^2\over 8 \pi^2x^2y^2}
\end{equation}
and it is easy to see that the sum of these two diagrams reproduces
the second term in the expansion (4.10). Further diagrams (Fig.9c etc.)
describe the perturbative solution for the Green function, i.e. of the
equation $\nabla^2G(x,y)=\delta^4(x-y)$ (leading to a power series in
$\rho^2$ rather than $g$ as discussed above). Contributions of these
diagrams to $G(x,y)$ reproduce the higher order terms (in $\rho^2$)
(4.10).
In order to reproduce the scalar propagator (4.10) we need only the
first four terms in the effective Lagrangean (4.7) ($\sim \rho^2$), the
higher operators ($\sim \rho^4$ etc) do not contribute.

The situation is more subtle for the $W$ propagator
\begin{equation}
\langle A_{\mu}(x)A_{\nu}(y) \exp -L_I(0)\rangle
\end{equation}
since the boson Green function in the instanton background does not
exist due to zero modes of the operator $\Box_{\mu\nu}^I$. In its place
we use therefore the constraint Green function satisfying the equation
\begin{equation}
\Box_{\mu\alpha}G_{\alpha\nu}(x,y)=\delta_{\mu\nu}-\sum f_{\mu}^{(k)}
\Phi_{\mu}^{0(k)}(y)
\end{equation}
where $\Phi_{\mu}^{0(k)}(y)$
are the zero modes and $f_{\mu}^{(k)}$ the constraints (see ref. [30]).
Therefore, starting from the order $\sim \rho^4$ the second term in
eq.(4.15) enters the game. It appears that the structure of the higher
terms in the effective Lagrangean (4.3) ($\sim \rho^4G_{\mu\nu}^2$ and
higher) should be correlated with eq.(4.15). As to the $\rho^2$ part of
the $W$ boson propagator, it could be verified to be reproduced by the
diagrams in Fig. 10, at least on the mass shell (see ref.[16-18]).

Finally, let us discuss the next-to-leading terms in the effective
Lagrangean (4.3) (with logarithmic accuracy). We start from the
operator $\rho^4(\bar\phi \phi)^2$. The coefficient in front of
this operator can be obtained by comparison of the four-point diagrams
in Fig.11 calculated in the instanton background with the analogous result in
the effective Lagrangean approach.(One should not take into account the
disconnected diagram in Fig.12a since it contributes only to the second
term $(\rho^2\bar{\phi}\phi)^2$ of the expansion of the operator
$exp{-2\pi^2\rho^2\bar{\phi}\phi}$,see Fig.12b). Let us start with the diagram
in fig.11a where we can use the explicit expressions for the Green
functions (4.10) and consider it at relatively large separations
$\rho^2\ll x_i^2\ll\mu^{-2}$.The logarithmic contribution to this diagram comes
from the region of large $z$ such as $\rho^2\ll z^2\ll \mu^{-2}$
where
$\mu$ is the normalization point of the effective Lagrangean serving as
IR-cutoff for the $z$ integration (we consider $\mu^2\ll m^2$).
At that large separations one can use the asymptotic expansion (4.10)
(for massless propagators since $x_i,z\ll \mu^{-1}$) keeping only the first
 three terms $\sim \rho^6,~ \rho^2,$ and
$\rho^4$. As we discussed above they correspond to the diagrams in Fig.
9 a,b, and c respectively. The corresponding diagrams for the
four-point Green function are shown in Fig. 13.
It is easy to see that only
the diagrams in Fig. 13a and b give contributions $\sim \rho^4\ln \rho^2$
(times the four tails corresponding to outgoing particles). After some
 combinatorics one obtains the
$\rho^4(\bar \phi\phi)^2$ term in the effective Lagrangean in the form
\begin{equation}
-3 \lambda \rho^4\pi^2(\bar \phi\phi)^2\ln (\rho^2\mu^2)
\end{equation}
where 3 is the sum of two 3/2's coming from Fig. 13a and 13b.

It is convenient to calculate the coefficients in front of effective
Lagrangean operators prior to the shift $\phi\rightarrow
\phi+ve/\sqrt{2}$ as thus one does not have to trace how the
combinations
$\phi+ve/\sqrt{2}$ arise in the effective Lagrangean approach.
In this case we have to take into account the negative vertex $-\lambda
v^2(\bar \phi\phi)$, it leads to a negative contribution
$-\lambda \rho^4
v^2(\bar \phi\phi)$. The corresponding coefficient is obtained from
a comparison of the two-point diagrams in Fig.14. Again, in order to
obtain the logarithmic part of the diagram in Fig.14a it is sufficient
to keep the first three terms of the asymptotic expansion (4.10). The
corresponding diagrams are shown in Fig. 15 and the result is
\begin{equation}
\pi^2\lambda \rho^4 v^2 (\bar\phi\phi)\ln (\rho^2\mu^2)
\end{equation}
with 1 being the sum of 1/4 coming from the diagram in Fig. 15a and
3/4 from Fig. 15b.(There are also diagrams similar to Fig. 13c-g which
do not give the logarithmic contribution).

It is instructive to demonstrate how the coefficient (4.16) can be
calculated (with logarithmic accuracy) directly within the effective
Lagrangean approach. Indeed, the diagrams in Fig.13 which we did
calculate are the perturbative diagrams describing the mixing of the
operators
\begin{eqnarray}
G_I~=&& -{2\pi^2i\over g} \rho_I^2~ {\cal T}r\{ u_0\bar u_I
\sigma_{\alpha} \bar \sigma_{\beta} u_I
\bar u_0 G_{\alpha \beta}\}\nonumber\\
H_I~=&& +~2\pi^2\rho_I^2\bar \phi\phi~
\end{eqnarray}
with the operator $H^2\sim \rho^4(\bar \phi\phi)^2$. (This type of mixing
is familiar, e.g. in the treatment of perturbative gluon corrections
to weak decays, see e.g. ref.[32]). We have
\begin{equation}
{1\over 2}(H_I)^{\mu_2^2}~ \times~
(H_I)^{\mu_2^2}~\rightarrow~
{1\over 2}(H_I)^{\mu_1^2}~\times~
(H_I)^{\mu_1^2}
-
{3\lambda \over 8\pi^2}(\ln{\mu_2^2\over \mu_1^2})H_I^2
\end{equation}
from the diagram in Fig. 13a and
\begin{equation}
{1\over 2}(G_I)^{\mu_2^2}\times~
(G_I)^
{\mu_2^2}~\rightarrow
{}~ {1\over 2}(G_I)^{\mu_1^2}\times~
(G_I)^{\mu_1^2}-~{3\lambda \over 8\pi^2}(\ln{\mu_2^2\over \mu_1^2})H_I^2
\end{equation}
from the diagram in Fig. 13b (here 1/2 is the combinatorics factor).
Note that we do not consider here the one-loop corrections to a single
operator $H_I$ (or $G_I$) since they correspond to the disconnected
diagrams of the Fig.12 type and hence have nothing to do with the
$\rho^4$ term of the effective Lagrangian.
Now, since at $\mu_2^2\sim \rho^{-2}$ there are no logarithmic
contributions to the coefficient in front of $\rho^4(\bar \phi \phi)^2$
we reobtain eq. (4.16) as a result of the evolution of the first two
operators $G_I$ and$H_I$ in the effective Lagrangean (4.3) from the UV-cutoff
$\mu_2^2=\rho^{-2}$ to the normalization point of the effective
Lagrangean $\mu_1=\mu$.

Similarly, since the constant ( in the non-shifted Lagrangean (3.2))
$\lambda v^2$ carries dimension there
will be the mixing
\begin{equation}
{1\over 2}(H_I)^{\mu_2^2}~ \times~
(H_I)^{\mu_2^2}~\rightarrow~
{1\over 2}(H_I)^{\mu_1^2}~\times~
(H_I)^{\mu_1^2}~
+~{\lambda \over 8}v^2\rho^2(\ln{\mu_2^2\over \mu_1^2})H_I
\end{equation}
coming from the diagram in Fig.15a and
\begin{equation}
{1\over 2}(G_I)^{\mu_2^2}\times~
(G_I)^
{\mu_2^2}~\rightarrow~ {1\over 2}(G_I)^{\mu_1^2}\times~
(G_I)^{\mu_1^2}+~{3\lambda \over 8}v^2\rho^2(\ln{\mu_2^2\over \mu_1^2})H_I
\end{equation}
coming from Fig. 15b. Again, taking $\mu_2^2=\rho^{-2}$ as the initial
point of the evolution we reobtain eq.(4.17).

Now let us turn to the coefficient in front of $(\bar{\phi}\phi)^2$
proportional to $g^2$ which comes from the diagram in Fig.11b. Again,
if we consider this diagram at large $x_i^2 \ss \rho^2$ (but $\ll
m^{-2}$) we can leave only the $\rho^0,\rho^2$,and $\rho^4$ terms in
the expansion of propagators in the instanton
background.Unfortunately,
the explicit form of the $\rho^4$ term in the expansion of the gluon
propagator is unknown(and depends on the constraint, see the
discussion
above).But with our accuracy we do not need it since the logarithmic
contribution comes only from the diagram in Fig.16a and the
corresponding term in the effective Lagrangian is
\begin{equation}
{3\over 8}g^2 \rho^4\pi^2(\bar \phi\phi)^2\ln (\rho^2\mu^2)
\end{equation}
It corresponds to the same type of mixing of the two gauge operators
with $\bar{\phi}\phi$ as in the eq.(4.20).

The method just desribed allows us also to determine the coefficients
in front of the remaining operators of order $\rho^4$, namely the
scalar-gauge operator $\sim \bar{\phi}G\phi$ and the gauge ones $\sim
G*G$, as a result of the evolution of the operators $G_i$ and$H_I$
from $\mu_2^2=\rho^{-2}$ to $\mu_1^2=\mu^2$. (There exists also the
operator
$(\nabla \bar{\phi})(\nabla\phi)$ of the same dimension which,
however, does not contribute to $U_{int}$ in the order $\epsilon^{8/3}$).
 After simple but somewhat
lengthy calculations one obtains the mixing in the form
\begin{equation}
{1\over 2}(G_I)^{\mu_2^2}\times~
(G_I)^
{\mu_2^2}~\rightarrow~
{1\over 2}(G_I)^{\mu_1^2}\times~
(G_I)^{\mu_1^2}~+~{g^2\over 4\pi^2}(\ln{\mu_2^2\over
\mu_1^2})(O_I+{5\over 4}P_I)
\end{equation}
and
\begin{eqnarray}
{1\over 2}(H_I)^{\mu_2^2}\times~
(G_I)^
{\mu_2^2}~\rightarrow~ \nonumber \\
{1\over 2}(H_I)^{\mu_1^2}\times~
(G_I)^{\mu_1^2}~-~{g^2\over 16\pi^2}(\ln{\mu_2^2\over \mu_1^2})P_I
\end{eqnarray}
where
\begin{eqnarray}
O_I=&&
{2\pi^4\rho^4\over g^2}(G_{\alpha\beta}^a
G_{\alpha\beta}^a
+2G_{\alpha\beta}^a
\tilde G_{\alpha\beta}^a\nonumber \\
&&-{\cal T}r\{u_0\bar{u}_I\sigma_{\alpha} \bar \sigma_{\zeta}u_I \bar{u}_0
G_{\beta\zeta}\}
{\cal T}r\{u_0\bar{u}_I\sigma_{\beta} \bar \sigma_{\eta}u_I \bar{u}_0
G_{\alpha\eta}\})\nonumber \\
P_I=&&
-{2\pi^4\rho^4\over g} i\bar{\phi}\{u_0\bar{u}_I\sigma_{\mu} \bar
\sigma_{\nu}u_I\bar{u}_0,G_{\mu\nu}\}\phi
\end{eqnarray}
We have used here the gauge-invariant external-field technique (see
e.g. ref.[32]). In terms of usual perturbative diagrams, the mixing
in eq.(4.24) with the gauge operator $O_I$ and the scalar-gauge
operator $P_I$ come from the diagrams in Figs.17 and 18
respectively and the mixing (4.25) is depicted in Fig.19. Again,
evolving the $\rho^2G$ and $\rho^2\bar{\phi}\phi$ operators
 from $\mu_2^2=\rho^{-2}$ (where the
coefficients in front of the gauge-operators contain no logarithmic
terms) to $\mu_1^2=\mu^2$ we obtain the corresponding contribution to
the effective Lagrangean in the form
\begin{equation}
{g^2\over 4\pi^2}(\ln{\rho^2 \mu^2})(O_I+P_I)
\end{equation}
To find the non-logarithmic term one should really calculate the
four-particle amplitudes in the instanton background (and use the exact
constrained Green function instead of the first few terms of its
asymptotic expansion) but at logarithmic accuracy we could avoid this
terrifying perspective.
The final form of the effective Lagrangean up to $\rho^4\ln\rho^2$
terms is
\begin{eqnarray}
L^I=&&
G_I~+~H_I~+~{g^2\over 4\pi^2}(\ln{\rho^2\mu^2})(O_I+P_I+{3\over 8}H_I^2)+
\nonumber \\
&&~{\lambda\over 4\pi^2}(\ln{\rho^2\mu^2})(-3H_I^2+2\pi^2v^2\rho^2 H_I)+~
 O(\rho^4 \cdot {\rm const})
\end{eqnarray}
In order to find  the corresponding Lagrangean after spontaneous
symmetry breaking we shift the fields according to
$\phi\rightarrow \phi+ve/\sqrt{2}$
($\bar\phi\rightarrow \bar\phi+v\bar e/\sqrt{2}$), and obtain
\begin{eqnarray}
L^I&&=
G_I+H_I+V_I+\pi^2v^2\rho^2+{g^2\over{4\pi^2}}
\ln{\rho^2\mu^2}(O_I+P_I+Q_+E_I+
\nonumber \\
&&{3\over 8}(H_I^2+2H_IV_I+V_I^2+
2\pi^2v^2\rho^2H_I+2\pi^2v^2\rho^2V_I+\pi^4v^4\rho^4)
\nonumber \\
&&+{\lambda\over
4\pi^2}\ln{\rho^2\mu^2}(-3H_I^2-6H_IV_I-3V_I^2
-4\pi^2v^2\rho^2H_I
\nonumber \\
&&-4\pi^2v^2\rho^2V_I-\pi^4v^4\rho^4
{}~+~ O(\rho^4 \cdot {\rm const})
\end{eqnarray}
where we used the notations
\begin{eqnarray}
V_I=&&
\sqrt{2}\pi^2\rho^2v(\bar{\phi}e+\bar{e}\phi) \nonumber\\
Q_I=&&
-{\sqrt{2}\pi^4\rho^4\over g} iv(\bar{\phi}\{u_0\bar{u}_I\sigma_{\mu} \bar
\sigma_{\nu}u_I\bar{u}_0,G_{\mu\nu}\}e
+\bar{e}\{u_0\bar{u}_I\sigma_{\mu} \bar
\sigma_{\nu}u_I\bar{u}_0,G_{\mu\nu}\}\phi) \nonumber\\
E_I=&&
-{\pi^4\rho^4\over g} iv^2\bar{e}\{u_0\bar{u}_I\sigma_{\mu} \bar
{\sigma}_{\nu}u_I\bar{u}_0,G_{\mu\nu}\}e
\end{eqnarray}
The answer for the antiinstanton effective Lagrangian $L^{\bar{I}}$
is obtained
by the substitution $u\bar{u}_0\rightarrow \bar{u}_{\bar I}$ and
$u_0 \bar{u}\rightarrow u_{\bar I}$ (which implies also changing of all
$\sigma$'s
into $\bar\sigma$'s and vice versa,e.g.$G_{\bar I}= -{2\pi^2i\over g}
\rho_{\bar I}^2~ {\cal T}r\{ u_{\bar I}\bar
\sigma_{\alpha}  \sigma_{\beta}
\bar{u}_{\bar I}G_{\alpha \beta}\}$).

In the next section we shall use this effective Lagrangean
to reproduce the valley result for the $\epsilon^{8/3}\ln{\epsilon}$
term of the holy grail function.

\vskip 1 cm
\setcounter{chapter}{5}
\setcounter{equation}{0}
\LARGE
5. Effective Lagrangean calculation of the holy grail function
\normalsize
\vskip 0.6 cm

In the effective Lagrangean approach the amplitude for forward $W$
scattering in the $I\bar I$ background (3.1) can be written as
\begin{eqnarray}
A(p,k)= \int {d\rho_1\over \rho_1^5}~d(\rho_1)~
\int {d\rho_2\over \rho_2^5}~d(\rho_2)~ \int dR~\int du~
\langle A_{\mu}^a(p) A_{\nu}^b(k) \nonumber \\
A_{\mu}^a(-p) A_{\nu}^b(-k)
L^I_{\psi}(0) \exp (-L_I(0))
L^{\bar I}_{\psi}(R) \exp (-L_{\bar I}(R)) \rangle
\end{eqnarray}
where $L_I$ and $L_{\bar I}$ are given in eq. (4.28) and we choose
$u_0=u_{\bar I}$ so $u\equiv u_I$ will be the matrix of relative
$I\bar{I}$ orientation.
The coefficients in front of the effective Lagrangean operators
which come from the integration over high momenta and can therefore be
calculated prior or after the shift $\phi \rightarrow
\phi+ve/\sqrt{2}$. In contrast the matrix elements of the effective
Lagrangean we are considering in this section are determined by the
region of small momenta
and they have to be calculated for the physical, massive theory, i.e.
after the shift
$\phi \rightarrow
\phi+ve/\sqrt{2}$, in order to avoid infrared divergences.
As discussed in section 1 we can neglect hard-hard and hard-soft
corrections at the level of accuracy we are interested in. Then,
$A_{\mu}(p)$ is given by the large-distance asymptotics of the sum of
$I$ and $\bar I$ fields (see eq.(3.5)) and the expression (5.1) reduces
to
\begin{equation}
A(p,k)\equiv  \int d\rho_1 d\rho_2 dR du ~d(\rho_1)
d(\rho_2)~\exp (iER_0)
\langle
\exp (-L_I(0))~
\exp (-L_{\bar I}(R)) \rangle
\end{equation}
with exponential accuracy. The BNV cross section is given by the
discontinuity of this amplitude continued to imaginary energies
(cf. eq.(3.7)) :
\begin{equation}
\sigma_{BNV}\equiv  {\cal I}m ~\int d\rho_1 d\rho_2 dR du
{}~d(\rho_1)d(\rho_2)~\exp (ER_0)
\langle
\exp (-L_I(0))~
\exp (-L_{\bar I}(R)) \rangle
\end{equation}
It is convenient to separate corrections to the instanton density
given by the disconnected contributions to the correlator in eq. (5.3)
from the $I\bar{I}$ interaction. We have
\begin{equation}
\langle
\exp (-L_I(0))~
\exp (-L_{\bar I}(R)) \rangle
= \exp \left( -S_H^I -S_H^{\bar I} + U_{\rm int}(\rho_1,\rho_2,
R,u) \right)
\end{equation}
where $S_H^I$ and $S_H^{\bar I}$
are the additional contributions to the instanton action $8\pi^2/g^2$
due to the Higgs condensate and $U_{\rm int}$ is the interaction
potential. Using the standard virial expansion we obtain
\begin{equation}
S_H^I = \langle L_I \rangle
-{1\over 2} \left( \langle L_I^2 \rangle
- \langle L_I \rangle^2 \right)
+{1\over 6} \left( \langle L_I^3 \rangle
- 3 \langle L_I^2 \rangle
\langle L_I \rangle^2 +2
\langle L_I \rangle^3 \right) + ...
\end{equation}
(similarly for $\bar I$) and
\begin{eqnarray}
U_{\rm int}&& = \langle L_I L_{\bar I}\rangle
-\langle L_I \rangle \langle L_{\bar I}\rangle
-{1\over 2} \left(
\langle L_I L_{\bar I}^2\rangle
+\langle L_I^2 L_{\bar I}\rangle
-\langle L_I \rangle \langle L_{\bar I}^2\rangle
-\langle L_I^2 \rangle \langle L_{\bar I}\rangle \right. \nonumber \\
&& \left. -2 \left[ \langle L_I \rangle +\langle L_{\bar I}\rangle\right]
\left[ \langle L_I L_{\bar I}\rangle
- \langle L_I \rangle \langle L_{\bar I}\rangle\right] \right)
+...
\end{eqnarray}
Since each $L_I$ contains at least one power of $\rho^2$ we
have to keep only the first few terms of the virial
expansions (5.5) and (5.6).

Let us start with the disconnected contributions of (5.4) corresponding
to the corrections to the instanton density $d(\rho)\sim e^{-8\pi^2
/g_W^2}$  coming from $S_H^I$ and $S_H^{\bar I}$  (see eq. (5.5)).
 The first term of the r.h.s.
of eq.(5.5) is obtained simply by inspection of eq.(4.28):
\begin{equation}
\langle L_I \rangle = \pi^2 v^2 \rho_1^2 - {\pi^2\lambda v^4\over 4} \rho_1^4
\ln (
\mu^2 \rho_1^2) +
{3\pi^2g^2 v^4\over 32} \rho_1^4 \ln (
\mu^2 \rho_1^2)
\end{equation}
The second and third term correspond to the diagrams in Fig. 20a
and 20b respectively. The logarithmic contributions come from
the loop momenta $\rho^{-2} \gg p^2 \gg \mu^2$. The second term on the
r.h.s. of eq. (5.5) is
\begin{eqnarray}
-{1\over 2} \left( \langle L_I^2 \rangle
- \langle L_I \rangle^2 \right)&=&
-{1\over 2} \langle V_I^2 \rangle
-{1\over 2} \langle G_I^2 \rangle
{}~=~ -2\pi^4 v^2 \rho_1^4 ~\int {dp\over (2\pi)^2} ~ {1\over p^2+m_H^2}
\nonumber \\
&&~ -{24 \pi^4\rho_1^4 \over g^2}~ \int
{dp\over (2\pi)^2} ~ {p^2\over p^2+m_W^2}
\end{eqnarray}
Subtracting, as usual, the quadratic (and higher) ultraviolet
divergencies, we obtain at the logarithmic accuracy we are interested in
\begin{equation}
-{1\over 2} \left( \langle L_I^2 \rangle
- \langle L_I \rangle^2 \right)=
{\pi^2 v^2 m_H^2 \rho_1^4\over 8}~ \ln {\mu^2 \over m_H^2} -
{3\pi^2 m_W^4 \rho_1^4\over 2 g^2}~ \ln {\mu^2 \over m_W^2}
\end{equation}
Combining eq.(5.7) and (5.9) we see that the normalization point $\mu$
drops out, as it should be, and the final result reads:
\begin{equation}
S_H^I= \pi^2 v^2 \rho_1^2 - {\pi^2 m_H^2 v^2\over 8} \rho_1^4 \ln (
m_H^2 \rho_1^2) +
{3\pi^2m_W^4\over 2g^2} \rho_1^4 \ln (
m_W^2 \rho_1^2)
\end{equation}
The higher terms in the expansion (5.5) contain extra powers of
$\rho^2m^2$ such that we can disregeard them.
$S_H^{\bar I}$ is obtained from $S_H^I$ by simply substituting
$\rho_1 \rightarrow \rho_2$.

It is worth noting that the logarithmic contribution to eq.(5.8)
obtained by expanding the propagators in powers of $m_H^2$ and
$m_W^2$ can be depicted by the same diagrams as those of Fig. 20,
in which case the vertices denote $m_H^2$ and $m_W^2$ mass
insertions. Then, the normalization point $\mu$ has to be interpreted
as boundary between the high momentum region, contributing to the
coefficient functions in front of the effective Lagrangean operators
(see eqs. (4.28) and (5.7)), and the low momentum region,
contributing to the matrix elements of these operators (see eq.(5.9)).

Let us turn now to the interaction between $I$ and $\bar I$. For our
accuracy the expansion (5.6) takes the form (there is also a
contribution $\sim \langle G_I^2(0) V_I(R) \rangle$ of order
$\epsilon^{8/3}$ but without $\ln \epsilon$) :
\begin{eqnarray}
U_{\rm int}&& = \langle G_I(0)G_{\bar I}(R) \rangle +
\langle V_I(0)V_{\bar I}(R) \rangle -
{1\over 2} \left( \langle G_I(0)^2G_{\bar I}(R) \rangle +
\langle G_I(0)G_{\bar I}^2(R) \rangle \right) \nonumber \\
&&- \left( \langle G_I(0)G_{\bar I}(R)V_{\bar I}(R) \rangle +
\langle G_I(0)V_I(0)G_{\bar I}(R) \rangle \right)
\nonumber \\
&&+{1\over 4} \left( \langle G_I^2(0)G_{\bar I}^2(R)\rangle -
\langle G_I^2(0)\rangle \langle G_{\bar I}^2(R) \rangle \right) \nonumber \\
&&+{g^2\over 4\pi^2} \ln (\rho_1^2\mu^2) ~\left( -{1\over 2}
\langle O_I(0)G_{\bar I}^2(R) \rangle
+\langle E_I(0)G_{\bar I}(R) \rangle \right) \nonumber \\
&&+{g^2\over 4\pi^2} \ln (\rho_2^2\mu^2) ~\left( -{1\over 2}
\langle G_I^2(0) O_{\bar I}(R)\rangle
+\langle G_I(0)E_{\bar I}(R) \rangle \right)
\end{eqnarray}
The first term corresponding to the diagram shown in Fig. 21a
is simply
\begin{eqnarray}
\langle G_I(0)G_{\bar I}(R) \rangle &&=
-{16\pi^4 \over g^2}~ \int {dq\over (2\pi)^4} e^{-iqR}
{4(q\cdot u)^2-q^2 \over q^2+m_W^2}
= {32\pi^2 \over g^2} {\rho_1^2\rho_2^2\over R^4}
\left( 4{(u\cdot R)^2\over R^2}-1 \right)\nonumber \\
&&\left\{ 1- {m_W^2R^2\over 8} +{m_W^4R^4\over 64} +
O(m_W^6R^6) \right\}
\end{eqnarray}
The first two terms in braces have the order $\epsilon^{4/3}$
and $\epsilon^2$ in the saddle point (1.2). (The first term is in fact
the dipole-dipole interaction (1.8).) The third term conteins no logarithmic
contribution
and therefore exceeds our accuracy. Similarly(see Fig. 21b),
\begin{eqnarray}
\langle V_I(0)V_{\bar I}(R) \rangle &&=
4\pi^4 v^4 \rho_1^2 \rho_2^2~ \int {dq\over (2\pi)^4} e^{-iqR}
{1\over q^2+m_H^2}\nonumber \\
&&= {\pi^2 v^2\rho_1^2\rho_2^2\over R^2}
\left( 1-{m_H^2R^2\over 4} \ln (m_H^2R^2) +
O(m_H^4R^4) \right) ~~~~~~
\end{eqnarray}
where the first term is $\sim \epsilon^2$
and the second $\epsilon^{8/3}\ln \epsilon$.

The third term on the r.h.s. of eq.(5.11) comes from the diagrams shown
in Fig. 22. The graph in Fig. 22a contributes
\begin{eqnarray}
{16\pi^2\over g^2} \rho_1^2\rho_2^2(\rho_1^2+\rho_2^2)~
\int {dq\over (2\pi)^4} e^{-iq\cdot R} {1\over q^2+m_W^2}
\int {dp\over (2\pi)^4} e^{-iq\cdot R}
{}~~~~~~~~~~\\
{16 (q\cdot u) \left( (p\cdot u)(q\cdot (2p-q)) -
(p\cdot(2p-q))(q\cdot u) \right) +8(p^2q^2-(p\cdot q)^2)
\over (p^2+m_W^2)((q-p)^2+m_W^2)}
\nonumber
\end{eqnarray}
Expanding the numerators in powers of $m_W^2$ we obtain
\begin{eqnarray}
{2\pi^4\over g^2} \rho_1^2\rho_2^2(\rho_1^2+\rho_2^2)~
\int {dq\over (2\pi)^4} e^{-iq\cdot R}
\left\{ {\Gamma(2-d/2)\over (q^2)^{2-d/2}} -{1\over 2-{d\over 2}} \right\}
\nonumber \\
\left( 4{(u\cdot R)^2\over q^2}-1 \right)
\left( q^2+5m_W^2 + O(m_W^4/q^2) \right)
\end{eqnarray}
where the second term in braces is the counterterm added in the
$\overline{MS}$ scheme (see e.g. ref.[29]). Thus
\begin{eqnarray}
-{1\over 2} \left( \langle G_I^2(0)G_{\bar I}(R) \rangle
+ \langle G_I(0)G_{\bar I}^2(R) \rangle\right) =
-{64\pi^2 \over g^2}~ \left(
{4(R\cdot u)^2 \over q^2}-1\right)
\nonumber \\
{\rho_1^2\rho_2^2(\rho_1^2+\rho_2^2)\over R^6}
\left\{ 1+ {5\over 16} m_W^2R^2 \ln(R^2\mu^2)+
O(m_W^2R^2) \right\}
\end{eqnarray}
where we have added the contribution of the diagram in Fig. 22b (due to the
commutator term in $G_{\mu\nu}$) which is
\begin{equation}
-{16\pi^2\over g^2}
{\rho_1^2\rho_2^2(\rho_1^2+\rho_2^2)\over R^6}
\left(
{4(R\cdot u)^2 \over q^2}-1\right)\left(1+
O(m_W^2R^2) \right)
\end{equation}
The factor 5/16 in eq.(5.17) is the sum of 3/8 coming from the mass
insertion shown in Fig. 23a and -1/16 coming from Fig. 23b. The first
term in braces in eq.(5.16) is of order $\epsilon^2$ and the second
of order $\epsilon^{8/3}\ln \epsilon$. Another term of this order
is
\begin{eqnarray}
- \langle G_I(0)G_{\bar I}(R)V_{\bar I}(R) \rangle
- \langle G_I(0)V_I(0)G_{\bar I}(R) \rangle =\nonumber \\
-v^2
{\rho_1^2\rho_2^2(\rho_1^2+\rho_2^2)\over R^4}
\left( {4(R\cdot u)^2 \over R^2}-1\right)
\ln(R^2\mu^2)
\end{eqnarray}
generated by Fig. 24.

Similarly to the case of $S_I^H$ considered above, the logarithmic
$\mu$ dependence of the matrix elements of the operator-correlators
from the r.h.s. of eq.(5.11) has to be cancelled by the $\ln \mu^2$
terms coming from the coefficient functions in front of the operators in the
effective Lagrangean. For example, for the two correlators just considered
the relevant operator is $(g^2/4\pi^2) \ln (\rho^2\mu^2)E$ and
we obtain (see Fig. 25)
\begin{eqnarray}
{g^2\over 4\pi^2} \ln (\mu^2\rho_1^2) \langle E_I(0)G_I(R)\rangle
+{g^2\over 4\pi^2} \ln (\mu^2\rho_2^2) \langle G_I(0)E_I(R)\rangle
\nonumber \\
={4\pi^2 \over g^2}~ \left(
{4(R\cdot u)^2 \over R^2}-1\right)
\rho_1^2\rho_2^2\left( \rho_1^2 \ln (\mu^2\rho_1^2)
+\rho_2^2 \ln (\mu^2\rho_2^2) \right)
\end{eqnarray}
Combining eqs.(5.17-19) gives the contribution to $U_{\rm int}$
in the form
\begin{eqnarray}
{64\pi^2 \over g^2}~ \left(
{4(R\cdot u)^2 \over R^2}-1\right)
\left\{ {\rho_1^2\rho_2^2(\rho_1^2+\rho_2^2)\over R^6} \right.
\nonumber \\
\left. +{m_W^2\rho_1^2\rho_2^2\over 4R^4}\left( \rho_1^2 \ln (R^2/\rho_1^2)
+ \rho_2^2 \ln (R^2/\rho_2^2) \right) \right\}
\end{eqnarray}
which in fact does not depend on $\mu$. Again, it is instructive
to note that we calculated both times the same diagrams  (Fig.
23a, 23b, and 24) with the loop momenta $\mu^2\gg p^2\gg R^2$
ascribed to matrix elements of correlators (eq. (5.17) and (5.18))
and momenta $\rho^{-2}\gg p^2 \gg \mu^2$ to the coefficient in front of the
opertaor
$E$ (=$P$, see diagrams in Fig. 18 and 19).

The last term of order $\epsilon^{8/3}\ln \epsilon$ is represented by the
 correlator
$\langle G_I^2(0)G_{\bar I}^2(R) \rangle$ (see Fig.26).The explicit
calculation of this correlator is rather tedious (see Ref.[19]) but with
logarithmic accuracy the answer can be easily restored from eq.(4.24)
since we know that the logarithmic dependence on $\mu$ in this
correlator should be canceled with the $\ln(\mu^2\rho^2)$ term in the
coefficient function in front of the operator $O_I$ in the effective
Lagrangian. The result is
\begin{eqnarray}
-{1\over 2} \left(  \langle G_I^2(0)G_{\bar I}^2(R) \rangle
- \langle G_I^2(0)\rangle \langle G_{\bar I}^2(R) \rangle \right)
{}~~~~~~~~~~~~~~~~~~~~ \\
- {64\pi^2 \over g^2}
{\rho_1^4\rho_2^4\over R^8} \left( 6+
2\left( {4(R\cdot u)^2 \over R^2}-1\right)^2-2\left( {4(R\cdot u)^2 \over
 R^2}-1\right)\right)
\ln(R^2\mu^2)
\nonumber
\end{eqnarray}
(it is worth noting that similar correlator but with $G_I^2(R)$
corresponding to the interaction of the two instantons vanishes as one
should expect from general considerations).
The result eq.(5.21) coincides with the calculation in ref.[19] for the $I$ and
$\bar{I}$ with maximal attractive orientations which only contribute to
$F(\epsilon)$,but for arbitrary orientations it
disagrees with the answer in ref.[19] unfortunately. We have (see Fig.27)
\begin{eqnarray}
-{g^2\over 8\pi^2} \ln (\mu^2\rho_1^2) \langle O_I(0)G_{\bar I}^2(R)\rangle
-{g^2\over 8\pi^2} \ln (\mu^2\rho_2^2) \langle G_I^2(0)O_{\bar I}(R)\rangle
{}~~~~~~~~~~~~~~ \\
= {64\pi^2 \over g^2}
{\rho_1^4\rho_2^4\over R^8}\left( 6+
2\left( {4(R\cdot u)^2 \over R^2}-1\right)^2-2\left( {4(R\cdot u)^2 \over
 R^2}-1\right)\right)
\ln(\rho_1\rho_2\mu^2)
\nonumber
\end{eqnarray}
So the contribution to $U_{\rm int}$ takes the form
\begin{equation}
- {64\pi^2 \over g^2}
{\rho_1^4\rho_2^4\over R^8} \left( 6+
2\left( {4(R\cdot u)^2 \over R^2}-1\right)^2-2\left( {4(R\cdot u)^2 \over
 R^2}-1\right)\right)
\ln{R^2\over \rho_1\rho_2}
\end{equation}
Note that by calculating the coefficient in front of the two-gluon
or ($O_I$) in the effective Lagrangean we, in fact, reproduced the result
obtained in ref.[19] by a hard two-loop calculation.

Thus the final form of $U_{\rm int}$ (up to $\epsilon^{8/3}\ln \epsilon$)
is
\begin{eqnarray}
U_{\rm int}&&=
{32\pi^2\over g^2}
{\rho_1^2\rho_2^2\over R^4}
\left(
{4(R\cdot u)^2 \over R^2}-1\right)
\nonumber \\
&&-{64\pi^2 \over g^2}~ \left(
{4(R\cdot u)^2 \over R^2}-1\right)
{\rho_1^2\rho_2^2\over R^4}
\left( {\rho_1^2+\rho_2^2\over R^2} + {m_W^2R^2\over 16}  \right)
\nonumber \\
&& +{\pi^2v^2\rho_1^2\rho_2^2\over R^2}
- {64\pi^2 \over g^2}
{\rho_1^4\rho_2^4\over R^8} \left( 6+
2\left( {4(R\cdot u)^2 \over R^2}-1\right)^2
\right. \nonumber \\
&&-2\left. \left( {4(R\cdot u)^2 \over
 R^2}-1\right)\right)
\ln{R^2\over \rho_1\rho_2}
\nonumber \\
&&
- {16\pi^2 \over g^2}
\left( {4(R\cdot u)^2 \over R^2}-1\right)
{m_W^2\rho_1^2\rho_2^2\over R^4}
\left( \rho_1^2\ln{R^2\over \rho_1^2}
+ \rho_2^2\ln{R^2\over \rho_2^2}\right)
\nonumber \\
&&-{\pi^2v^2\over4} m_H^2\rho_1^2\rho_2^2 \ln(m_H^2R^2)
\end{eqnarray}
The first two terms in $U_{\rm int}$ for the pure gauge sector
reproduce the first two terms of the expansion of the conformal
expression (2.31) but the third deviates from it. This means that
a calculation of $U_{\rm int}$ using the effective Lagrangean corresponds to
using a different valley than that in eq.(2.26). However, the final
result for the holy grail function $F(\epsilon)$ is the same in both
cases. Indeed we have
(see eq. (5.4))
\begin{eqnarray}
-S_I^H-S_{\bar I}^H+U_{\rm int}=
-\pi^2v^2(\rho_1^2+\rho_2^2)
+{\pi^2v^2\rho_1^2\rho_2^2\over R^2}
{}~~~~~~~~~~~~~~~~~~~~~~~~~~~~~~~~~~~~~~~~~\nonumber \\
+{32\pi^2\over g^2}
{\rho_1^2\rho_2^2\over R^4}
\left(
{4(R\cdot u)^2 \over R^2}-1\right)
-{64\pi^2 \over g^2}~ \left(
{4(R\cdot u)^2 \over R^2}-1\right)
{\rho_1^2\rho_2^2\over R^4}
\left( {\rho_1^2+\rho_2^2\over R^2} + {m_W^2R^2\over 16}  \right)
\nonumber \\
 -{\pi^2v^2m_H^2\over 8}
\left( \rho_1^4 \ln(m_H^2\rho_1^2)
+  \rho_2^4 \ln(m_H^2\rho_2^2)\right)
+{3\pi^2m_W^4\over 2g^2}
\left( \rho_1^4 \ln(m_W^2\rho_1^2)
+  \rho_2^4 \ln(m_W^2\rho_2^2)\right)
\nonumber \\
- {64\pi^2 \over g^2}
{\rho_1^4\rho_2^4\over R^8} \left(  6+
2\left( {4(R\cdot u)^2 \over R^2}-1\right)^2-2\left( {4(R\cdot u)^2 \over
 R^2}-1\right)\right)
\ln{R^2\over \rho_1\rho_2} ~~~~~~~~~~~
\nonumber \\
- {16\pi^2 \over g^2}
\left( {4(R\cdot u)^2 \over R^2}-1\right)
{m_W^2\rho_1^2\rho_2^2\over R^4}
\left( \rho_1^2\ln{R^2\over \rho_1^2}
+ \rho_2^2\ln{R^2\over \rho_2^2}\right) ~~~~~~~~~~~~~~~~~~~~~
\nonumber \\
-{\pi^2v^2\over4} m_H^2\rho_1^2\rho_2^2 \ln(m_H^2R^2)
{}~~~~~~~~~~~~~~~~~~~~~~~~~
\end{eqnarray}
Evaluating this expression at the saddle   point (1.2) one reproduces the
valley result (1.5) although the explicite form of equ.(5.24) and (3.45)
differ.

\newpage
\setcounter{chapter}{6}
\setcounter{equation}{0}
\LARGE
6. Conclusions
\normalsize
\vskip 0.6 cm

We have calculated the $\epsilon^{8/3}\ln \epsilon$ term of the holy
grail function with two methods, namely the valley method and the
effective Lagrangean approach.
Though the final result for both methods
is the same they differ completely in the way it is obtained.
The effective Lagrangean approach is more pictorial and
also it gives us an opportunity to deal with the multi-instanton
amplitudes in a simple way: just expand several times in powers of $
L_I^{eff}$ and $ L_{\bar{I}}^{eff}$ and calculate the obtained
Feynman diagrams with additional multi-W (and multi-Higgs) vertices.
On the other hand , the valley method enables us to use the (tree-level)
conformal invariance in a pure gauge sector which gives the expansion
of $U^g_{int}$ in powers of conformal parameter $1/\xi=\rho_1\rho_2/
(R^2+\rho_1^2+\rho_2^2)$ instead of reconstructing it from the
expansion in powers of $\rho_1^2/R^2$ and $\rho_2^2/R^2$.Also, the valley
method saves us
from calculating the two-loop diagrams in Fig.26
(at a price of more complex contributions in the gauge-Higgs sector).

Of course, the
question one really would like to answer is how
these
instanton-induced cross sections
behave at SSC energies. Unfortunately, as we
mentioned above, in order to answer this question we have to continue
the expansion of $F(\epsilon)$ in $\epsilon$ which implies that we
must take into account not only
the $I\bar{I}$ potential $U_{int}$ but the hard-hard and hard-soft
quantum corrections as well.Also, this should be done in
the Minkowski space due to the reasons discussed in the Introduction.
The continuation of the effective-Lagrangian approach to the Minkowski
space is quite direct - one simply should take care of  the trivial
$i$'s and signs according to the general rules of Wick rotation.
(In fact, it is the most simple way to understand the instanton-induced
amplitudes in the Minkowski space). The valley method can also be
modified to meet our purposes. One can write down the functional
integral directly for the cross sections with BNV in the final state
(at a price of doubling of the number of fields).The valleys for this
double-set functional integral determine the cross sections with BNV
in the leading semiclassical approximation,see ref.[32]. However,
as we mentioned
above, at large energies the quantum corrections are also essential
and one faces the problem of determining the high-energy behavior
of the propagators in the background of these valley fields.The
similar problem within the effective-Lagrangian approach corresponds
to the summation of the diagrams of the type shown in Fig.10 but with
both $I$ and $\bar{I}$ effective vertices taken into account.
We hope to return to these questions in further publications.

\vskip 1 cm
Acknowledgement:

The authors are grateful to D.I.Diakonov for valuable discussions.One
of us (I.B.) would like to thank the Department of Theoretical Physics
at
Frankfurt University where part of this work was done for hospitality.
This work was supported in part by the US Department of Energy under the
grant DE-F902-90ER-40577 and in part by the Deutsche
Forschungsgemeinschaft (Scha 458/3-1).

\vskip 5 cm
REFERENCES :

\vskip 0.6 cm
[1]   G. t'Hooft, Phys.Rev. D14(1976)3432

\vskip 0.6 cm
[2]   A. Ringwald, Nucl.Phys. B330(1990)1;O. Espinosa, Nucl.Phys. B343(1990)310

\vskip 0.6 cm
[3]  V.V. Khoze and A. Ringwald, Phys.Lett. B259(1991)106

\vskip 0.6 cm
[4] V.V. Khoze, J. Kripfganz, and A. Ringwald, Phys.Lett.
 B 275(1992)381,

{}~~~~Erratum B279(1992)429; ibid B277(1992)496

\vskip 0.6 cm
[5]  T. Banks, G. Farrar, M. Dine, D. Karabali, and B. Sakita,

{}~~~~Nucl.Phys. B347(1990)581;

{}~~~~R. Singleton, L. Susskind, and L. Thorlaciuus, Nucl.Phys. B343(1990)541

\vskip 0.6 cm
[6] V.I. Zakharov, Nucl.Phys. B353(1991)683

\vskip 0.6 cm
[7] M. Maggiore and M. Shifman, Nucl.Phys. B371(1992)177

\vskip 0.6 cm
[8] C.G. Callan, R. Dashen, and D.J. Gross, Phys.Rev. D17(1978)271

\vskip 0.6 cm
[9] S. Khlebnikov, V. Rubakov, and P. Tinyakov, Nucl.Phys. B350(1991)441

\vskip 0.6 cm
[10] V.V. Khoze and A. Ringwald, Nucl.Phys. B355(1991)351

\vskip 0.6 cm
[11] V.I. Zhakarov, preprint TPI-MINN-90/7-T(1990), Nucl.Phys. B371(1992)637

\vskip 0.6 cm
[12] A.V. Yung, 'Instanton-induced effective Lagrangean in the Gauge-

{}~~~~Higgs
theory', preprint SISSA 181/90/EP (1990)

\vskip 0.6 cm
[13] I.I. Balitsky and A.V. Yung, Phys.Lett. B168(1986)113;

{}~~~~Nucl.Phys. B274(1986)475

\vskip 0.6 cm
[14] A.V. Yung, Nucl.Phys. B297(1988)47

\vskip 0.6 cm
[15] V.V. Khoze and A. Ringwald, 'Valley trajectories in gauge theories',

{}~~~~preprint
CERN-TH-60282/91 (1991)

\vskip 0.6 cm
[16] P. Arnold and M. Mattis, Phys.Rev.Lett 66(1991)13

\vskip 0.6 cm
[17] D.I. Diakonov and V.Yu. Petrov, Baryon number non-conservation

{}~~~~at high energy', proceedings of the XXVI LNPI Winter School,

{}~~~~Leningrad
1991

\vskip 0.6 cm
[18] A.H. Mueller, Nucl.Phys. B364(1991)109

\vskip 0.6 cm
[19] D.I. Diakonov and M. Polyakov, 'Baryon number non-conservation

{}~~~~at high energies
and instanton interactions',

{}~~~~preprint LNPI-1737(1991),

\vskip 0.6 cm
[20]
A.H. Mueller, Nucl.Phys. B348(1991)310, ibid B353(1991)44???

\vskip 0.6 cm
[21]
A.H. Mueller, Nucl.Phys. B381(1992)597

\vskip 0.6 cm
[22] I.I. Balitsky and V.M. Braun, Nucl.Phys. B380(1992)51

\vskip 0.6 cm
[23] E.B. Bogomolny, Phys.Lett. B91(1980)431;

{}~~~~ J. Zinn-Justin, Nucl.Phys. B192(1981)125

\vskip 0.6 cm
[24] I.I. Balitsky, Phys.Lett. B273(1991)282

\vskip 0.6 cm
[25] J. Verbaarshot, Nucl.Phys. B362(1991)33

\vskip 0.6 cm
[26]  N. Andrei and D.J. Gross, Phys.Rev. D18(1978)468

\vskip 0.6 cm
[27] L.S. Brown, R.D. Carlitz, D.B. Creamer , C. Lee,

{}~~~~Phys.Rev. D17 (1978)1583

\vskip 0.6 cm
[28] I. Affleck, Nucl.Phys. B191(1981)429

\vskip 0.6 cm
[29] J.C. Collins 'Renormalization', Cambridge University Press, 1984

\vskip 0.6 cm
[30] H. Levine and L. Yaffe, Phys.Rev. D19(1979)1225

\vskip 0.6 cm
[31] M.A. Shifman, A.I. Vainshtein, and V.I. Zakharov, Phys.Rev. D18(1978)2583

\vskip 0.6 cm
[32] I.I. Balitsky and V.M. Braun, Nucl.Phys. B311(1988/89)541

\newpage
\begin{center}{\bf Figure~ Captions~ :}
\end{center}

\noindent
\vskip 0.6 cm
Fig.1: Perturbative diagrams reproducing the classical instanton fiels ($a$)
and the propagator in the instanton background ($b$) in the
effective-Lagrangian approach. Small open circle denotes the
instanton-induced effective vertex $\exp\left({2\pi^2i\over g} \rho^2
 ~Tr\{ \sigma_{\alpha}\bar \sigma_{\beta}~G_{\alpha \beta}(x)\}\right )$.

\vskip 0.6 cm
\noindent
Fig.:2: Perturbative diagrams for $I\bar{I}$ interaction in the
effective-Lagrangian approach. Small full circle denotes the
antiinstanton effective vertex

$ \exp\left(
{2\pi^2i\over g} \rho^2~ Tr\{ u \bar \sigma_{\alpha}
\sigma_{\beta}\bar u ~G_{\alpha \beta}(x)\}\right)$.

\vskip 0.6 cm
\noindent
Fig.3: The $\epsilon^2$ contributions to the instanton-antiinstanton
interaction due to an additional W (Fig.1a), mass insertions (Fig.1b)
 and Higgs exchange (Fig.1c).

\vskip 0.6 cm
\noindent
Fig.4: Two examples of $\epsilon^{8/3}\ln\epsilon$ contributions to the
instanton-antiinstanton interaction.

\vskip 0.6 cm
\noindent
Fig.5: Two discontinuites of the hard-hard correction
corresponding to
the cross section with BNV ($a$) and without ($b$).

\vskip 0.6 cm
\noindent
Fig.6: Illustration of the valley configuration.

\vskip 0.6 cm
\noindent
Fig.7: Large-distance expansion of the instanton field in eq. (4.5) in terms of
 perturbative
diagrams induced by effective Lagrangian.

\vskip 0.6 cm
\noindent
Fig.8: Expansion of the scalar component of the instanton. A cross on
the end of a scalar line denotes the Higgs condensate.

\vskip 0.6 cm
\noindent
Fig.9: Scalar propagator in the
instanton background as a sum of the perturbative diagrams in the
effective-Lagrangian approach.

\vskip 0.6 cm
\noindent
Fig.10: Perturbative diagrams for the  W boson propagator in the field of an
instanton.

\vskip 0.6 cm
\noindent
Fig.11: Quartic scalar Green function in the instanton background -
connected diagrams.

\vskip 0.6 cm
\noindent
Fig.12: Disconnected part of the quartic Green function coming from the
square of scalar propagator in the instanton background.

\vskip 0.6 cm
\noindent
Fig.13: Quartic scalar Green function in the
effective-Lagrangian approach ($a$ and $b$ diagrams represent the
logarithmic mixing of the
operators  $H_I$ and $G_I$ with the four-Higgs
operator $\rho^4(\bar{\phi}\phi)^2$)

\vskip 0.6 cm
\noindent
Fig.14: Additional negative contribution to the Green function
of the non-shifted scalar field in the instanton background due to the
vertex $-\lambda v^2\bar{\phi}\phi$.

\vskip 0.6 cm
\noindent
Fig.15: The same additional term in the effective-Lagrangian approach
(only the logerithmic diagrams corresponding to mixing (4.21) and
(4.22) are depicted).

\vskip 0.6 cm
\noindent
Fig.16: The leading logarithmic perturbative diagram for the
 contribution of
the type of Fig.11b to the quartic scalar Green function in the
effective-Lagrangian approach (it corresponds to the
mixing of $G_I$ with the four-Higgs operator).

\vskip 0.6 cm
\noindent
Fig.17: One-loop diagrams for mixing of the operator $G_I$ with
the two-gluon operators $\sim \rho^4G*G$.

\vskip 0.6 cm
\noindent
Fig.18: Mixing of $G_I$ with the scalar-gluon operator $\sim
\rho^4 \bar{\phi}G\phi$
\vskip 0.6 cm
\noindent
Fig.19: Diagram for the mixing (4.25).

\vskip 0.6 cm
\noindent
Fig.20: Leading diagrams for the corrections to the instanton density
$\sim m_H^2v^2\rho^4$ ($a$) and $\sim {1\over g_W^2}m_W^4\rho^4$ ($b$)
in the
effective-Lagrangian approach.

\vskip 0.6 cm
\noindent
Fig.21: The leading contributions  to $I\bar{I}$
interaction due to exchange by W or Higgs boson given by the correlators
$\langle G_I(0)G_{\bar I}(R) \rangle$ ($a$) and $\langle V_I(0)V_{\bar
I}(R) \rangle$ ($b$).

\vskip 0.6 cm
\noindent
Fig.22: The next-to-leading graphs given by the correlator $\langle
 G_I(0)^2G_{\bar I}(R) \rangle$.

\vskip 0.6 cm
\noindent
Fig.23: The logarithmic contributions to $U_{int}$  $\sim
\epsilon^{8/3}\ln{\epsilon}$ coming from expanding the propagators in
Fig. 22 in powers of $m_W$.

\vskip 0.6 cm
\noindent
Fig.24: The  $\sim
\epsilon^{8/3}\ln{\epsilon}$ contribution to $U_{int}$ coming from the
correlator $\langle G_I(0)V_I(0)G_{\bar I}(R) \rangle$.

\vskip 0.6 cm
\noindent
Fig.25 The diagram for the correlator $\langle E_I(0)G_{\bar I}(R)
\rangle$. (The coefficient ${g^2\over 4\pi^2} \ln (\rho_2^2\mu^2)$ in
front of this correlator corresponds to the region of large momenta
$\rho^{-2}\gg p^2\gg \mu^2$ in the Feynman graphs shown in Figs. 23
and 24).

\vskip 0.6 cm
\noindent
Fig.26: The two-loop diagrams for the ${\pi^2 \over g^2}
{\rho_1^4\rho_2^4\over R^8}
\ln({R^2\over \rho_1\rho_2})$ part of $U_{int}$ given by the correlator
$\langle
 G_I^2(0)G_{\bar I}^2(R) \rangle$.

\vskip 0.6 cm
\noindent
Fig.27 The first-order diagrams  for the correlators $ \langle O_I(0)G_{\bar
I}^2(R)\rangle$ and $\langle G_I^2(0)O_{\bar I}(R)\rangle$.
\end{document}